\documentclass[aps,prd,longbibliography,noeprint,onecolumn,nofootinbib,showkeys,notitlepage,superscriptaddress]{revtex4-2} 
\usepackage{natbib}
\usepackage{graphicx}
\usepackage{dcolumn}
\usepackage{bm}
\usepackage{amsmath}
\usepackage{amssymb}
\usepackage{mathrsfs} 
\usepackage{float}
\usepackage{multirow}
\usepackage{threeparttable}
\usepackage{slashed}
\usepackage{xcolor}
\usepackage{braket}
\usepackage{multirow}
\usepackage{gensymb}
\usepackage[colorlinks=true, pdfstartview=Fit, bookmarks=true, bookmarksnumbered=true, breaklinks]{hyperref}
\usepackage{mathtools,braket}

\usepackage{color}
\definecolor{blue}{rgb}{0.0, 0.0, 1.0}
\definecolor{red}{rgb}{1.0, 0.0, 0.0}
\definecolor{royalblue}{rgb}{0.0, 0.14, 0.4}
\hypersetup{linkcolor=black,citecolor=blue,urlcolor=blue,pdftitle={Comparison of five potential models for semileptonic B-meson decays in the fully Bakamjian-Thomas-based light-front quark model},pdfauthor={Ho-Meoyng Choi, Bhoomika Pandya}}
\graphicspath{{figures/}{./}}
\usepackage{etoolbox}
\makeatletter
\AtBeginDocument{%
  \patchcmd{\@makecaption}{\small\rmfamily}{%
    \ifnum\csname ftype@\@captype\endcsname=1\normalsize\else\small\fi\rmfamily
  }{}{\PackageWarning{BtoDFigures}{Caption-font patch not applied}}%
}
\makeatother

\newcommand{\RDRange}{0.282\text{--}0.301}

\newcommand{\RDsRange}{0.287\text{--}0.303}

\newcommand{\RetacRange}{0.292\text{--}0.328}

\newcommand{\AuxMaxError}{0.52\%}

\newcommand{\nc}{\newcommand}
\nc{\non}{\nonumber}
\nc{\hc}{\hbox {H.c.}}
\nc{\noi}{\noindent}
\nc{\barx}{\bar{x}}
\nc{\pbarn}{\;\hbox {pb}}
\nc{\fbarn}{\;\hbox {fb}}
\nc{\cone}{{\scriptsize \textcolor{orange}{\circled{1}}}}
\nc{\ctwo}{{\scriptsize \textcolor{orange}{\circled{2}}}}
\nc{\cthr}{{\scriptsize \textcolor{orange}{\circled{3}}}}

\nc{\hsp}{\hspace{0.5cm}}
\nc{\lsp}{\hspace{1cm}}
\nc{\Lsp}{\hspace{2cm}}
\nc{\LLsp}{\lsp\lsp}
\nc{\lra}{\longrightarrow}
\nc{\p}{\prime}
\nc{\sgn}{\text{sgn}}
\nc{\ph}{\varphi}
\nc{\eff}{\mathrm{eff}}
\nc{\sqM}{\sqrt{M}}
\nc{\NL}{\mathrm{NL}}
\nc{\moc}{\mathcal{M}} 
\def\ShowRevisions{}
\ifdefined\ShowRevisions
  
\else
  
\fi
\nc{\bl}[1]{\textcolor{blue}{#1}}

\nc{\pbar}{\bar{\psi}}
\nc{\vp}[1]{\mathbf{p}_{#1}}
\nc{\vk}{{\bm k}_\perp}
\nc{\vkp}{{\bm k}'_\perp}
\nc{\vq}{{\bm q}_\perp}
\nc{\uu}{{\uparrow\uparrow}}
\nc{\ud}{{\uparrow\downarrow}}
\nc{\du}{{\downarrow\uparrow}}

\def\lsim{\mathrel{\raise.3ex\hbox{$<$\kern-.75em\lower1ex\hbox{$\sim$}}}}
\def\gsim{\mathrel{\raise.3ex\hbox{$>$\kern-.75em\lower1ex\hbox{$\sim$}}}}

\def\vp{{\bf p}}

\def\be{\begin{equation}}
\def\ee{\end{equation}}
\def\bea{\begin{eqnarray}}
\def\eea{\end{eqnarray}}

\def\vp{{\bf p}}

\def\be{\begin{equation}}
\def\ee{\end{equation}}
\def\bea{\begin{eqnarray}}
\def\eea{\end{eqnarray}}

\begin{document}

\title{A comparative analysis of semileptonic
$B_{(s)}\to D_{(s)}$ and $B_c\to(\eta_c,B,B_s)$ decays
in the light-front quark model}
\author{Ho-Meoyng Choi}
\email{homyoung@knu.ac.kr}
\affiliation{Department of Physics Education, Teachers College, Kyungpook National University, Daegu 41566, Republic of Korea}
\author{Bhoomika Pandya}
\affiliation{Instituto de F\'isica Corpuscular (IFIC),
CSIC--Universitat de Val\`encia, E-46980 Paterna, Valencia, Spain}

\begin{abstract}
We present a global analysis of the semileptonic transitions
$B_{(s)}\to D_{(s)}$ and $B_c\to(\eta_c,B,B_s)$ by comparing five
potential-model implementations of the QCD-motivated light-front
quark model. We use the same fully Bakamjian--Thomas-based
prescription for the weak current in all five models and keep
their respective spectroscopic parameters fixed.
We examine how confinement, radial mixing, and the treatment
of hyperfine interactions jointly affect the form factors,
differential decay rates, branching fractions, and lepton-flavor
ratios.
The internal Lorentz prefactors use the same free quark--antiquark
kinematics as the on-shell constituent spinors in the weak-current
kernels, while the physical meson masses determine the external
kinematics.
For each model and decay channel, the two independent extractions
of $f_-(q^2)$ using the $(+,\perp)$ and $(+,-)$ current-component
combinations agree within numerical accuracy.
We perform a direct analytic continuation of the spacelike overlap
integrals in the $q^+=0$ frame using a mass-weighted contour in
complex transverse momentum space. This yields the form factors
throughout the physical timelike region, including zero recoil,
without parametrizing their $q^2$ dependence.
Form factors and differential decay rates are presented as
equal-weight model averages with pointwise minimum--maximum bands.
The light-lepton-averaged ratios range from $0.282$ to $0.301$ for
$R(D)$ and from $0.287$ to $0.303$ for $R(D_s)$.
\end{abstract}

\maketitle
\vspace{-4pt}
\section{Introduction}

Semileptonic transitions between heavy pseudoscalar mesons provide
clean probes of weak interactions and nonperturbative QCD in
heavy--light and heavy--heavy systems. The $b\to c$ transitions
involve a $u$ or $d$ spectator in $B\to D$, a strange spectator
in $B_s\to D_s$, and a charm spectator in $B_c\to\eta_c$.
The $c\to d$ transition in $B_c\to B$ and the $c\to s$ transition
in $B_c\to B_s$ both have a bottom spectator. The hadronic matrix
elements are described by two independent form factors, which may
be chosen as the vector and scalar form factors $f_+(q^2)$ and
$f_0(q^2)$ or, equivalently, as $f_+(q^2)$ and
$f_-(q^2)$~\cite{Richman1995,Neubert1993,Korner1989}. These form
factors provide the hadronic input for extracting
Cabibbo--Kobayashi--Maskawa (CKM) matrix elements and testing
lepton-flavor universality.

The recoil spectra in $B\to D\ell\nu_\ell$ with $\ell=e,\mu$
primarily constrain the shape of $f_+(q^2)$.
BaBar analyzed the recoil and lepton-helicity-angle distributions
in $B\to D\ell\nu_\ell$ and determined the form factors and $|V_{cb}|$
by combining these data with lattice QCD results and earlier Belle
measurements~\cite{BaBar2024}.
Belle II combined its recoil spectra with lattice QCD form factors
to determine $|V_{cb}|$~\cite{BelleII2025Vcb}. LHCb measured the branching
fraction of $B_s^0\to D_s^-\mu^+\nu_\mu$ relative to that of
$B^0\to D^-\mu^+\nu_\mu$~\cite{LHCb2020}. The common
$|V_{cb}|^2$ factor cancels in this ratio, making it useful
for comparing the two spectator flavors. Using the LHCb
distribution of the $D_s$ momentum perpendicular to the $B_s$
flight direction, Earnshaw \textit{et al.} extracted $|V_{cb}|$
and constrained the shape of the $B_s\to D_s$ form
factor~\cite{Earnshaw2026}.

Complementary information comes from
$R(D)\equiv\mathcal B(B\to D\tau\nu_\tau)/
\mathcal B(B\to D\ell\nu_\ell)$. The scalar form factor $f_0$
contributes appreciably to the $\tau$ rate, whereas its contribution
to the light-lepton rates is helicity suppressed~\cite{Korner1989}.
BaBar and Belle measured $R(D)=0.440\pm0.072$ and
$0.307\pm0.040$, respectively~\cite{BaBar2013RD,Belle2020RD}.
For the charged-$D$ channel, Belle II obtained
$R(D^+)=0.418^{+0.090}_{-0.092}$ using electron and muon
normalization modes, while LHCb obtained $0.249\pm0.064$
using the muon mode~\cite{Belle2025,LHCb2024}.
The ICHEP 2026 average from the
Heavy Flavor Averaging Group (HFLAV), $R(D)=0.359\pm0.024$,
lies about $2.6$ standard deviations above the arithmetic average
of Standard Model predictions, $0.296\pm0.004$~\cite{HFLAV2026}.
Describing both the recoil spectra in the light-lepton modes
and $R(D)$ requires predictions for $f_+(q^2)$ and $f_0(q^2)$
over the physical decay region.

Lattice QCD provides first-principles determinations of these
form factors~\cite{Flag2024}.
For $B\to D$, HPQCD and Fermilab/MILC calculated $f_+$ and
$f_0$ at nonzero recoil, providing inputs for extracting $|V_{cb}|$
and predicting $R(D)$~\cite{HPQCD2015,FNALMILC2015}.
For $B_s\to D_s$, HPQCD determined both form factors and extended
them over the physical $q^2$ range using a $z$ expansion~\cite{HPQCD2017}.
A subsequent HPQCD calculation covered the full physical $q^2$
range using nonperturbatively normalized currents~\cite{McLean2020}.

Complementary studies of $B_{(s)}\to D_{(s)}$ transitions employ
perturbative QCD~\cite{Hu2019,Fan2013},
QCD sum rules~\cite{Blasi1993},
light-cone sum rules~\cite{Zhang2021,Gao2022},
relativistic quark models~\cite{Faustov2012,Patnaik2025},
and the covariant confined quark model~\cite{Soni2021,Soni2023,Ivanov2017}.
Heavy-quark expansions~\cite{Bordone2019,Bordone2025} and
dispersive analyses~\cite{Cohen2019,Gubernari2026} provide
additional constraints on the form factors and their recoil dependence.
For $B_c\to\eta_c$, calculations include covariant light-front
models~\cite{WSL09}, a symmetry-preserving continuum approach
based on Dyson--Schwinger and Bethe--Salpeter (BS)
equations~\cite{YaoBc2021}, light-cone sum rules~\cite{Zeng2025},
and basis light-front quantization~\cite{Tang2021}.
Semileptonic and nonleptonic $B_c$ decays to charmonium
have also been analyzed~\cite{Biswas2025}.
For the charm-changing decays $B_c\to B_{(s)}$,
HPQCD determined the form factors in lattice QCD over their
full physical momentum-transfer ranges~\cite{Cooper2020}.

Among constituent approaches, the light-front quark model
(LFQM), based on light-front (LF) dynamics~\cite{LB80,Brodsky1997},
provides a common framework for studying meson transitions
with different constituent-mass hierarchies.
Light-front wave functions (LFWFs) depend on internal variables
invariant under kinematic LF boosts, separating the
overall meson motion from internal constituent motion.
In the one-body current approximation, transition matrix elements
can be expressed as LFWF overlaps with relativistic spin
couplings~\cite{Jaus1990,Jaus1991}.
Both standard and covariant LF formulations have been
applied to heavy-meson weak
decays~\cite{Jaus1990,Jaus1999,CCH04,WSL09}.

Our earlier LFQM studies applied this framework to exclusive
pseudoscalar-to-pseudoscalar transitions~\cite{Choi1999}
and to semileptonic and radiative decays of the $B_c$
system~\cite{Choi2009}.
Extracting form factors requires consistent treatment of the
LFWFs, the constituent current, and the Lorentz projection,
particularly for transverse or minus current components.
Early studies examined LF zero-mode contributions and their
dependence on the current component and meson-vertex
structure~\cite{LFzeroCJ,BCJ01,LFzero05}.

The self-consistent type-II correspondence between the covariant
BS model and the standard LFQM was introduced in a study of
vector-meson decay constants~\cite{CJ14}.
This correspondence maps the on-shell contributions of the
covariant BS amplitudes to the LFQM by matching LF vertex
functions to LFQM wave functions and applying $M\to M_0$
to all $M$-dependent terms in the decay-constant integrands.
Here $M$ is the physical meson mass, and $M_0$ is the free
invariant mass of the on-shell constituent quark--antiquark pair.
Reference~\cite{Arifi2022} subsequently connected this prescription
to the Bakamjian--Thomas (BT) construction and demonstrated
current-component, polarization, and frame independence of
pseudoscalar and vector meson decay constants evaluated with
one-body currents in the LFQM.
In the LF realization of the BT construction, interactions enter
the invariant mass operator under constraints that preserve the
Poincar\'e algebra, while the kinematic generators retain their
free-particle form~\cite{BT53,Poly10,CCKP}.
Meson states in the LFQM are expanded in an on-shell constituent
quark--antiquark basis, with binding dynamics encoded in the
radial LFWFs.

For pseudoscalar-to-pseudoscalar semileptonic transitions,
Ref.~\cite{Choi2021} had already applied the same correspondence
to $B\to D\ell\nu_\ell$, providing a direct precursor to the
present work.
The replacements $M\to M_0$ and $M'\to M_0'$ were applied
to the initial and final mesons, respectively, throughout the
internal form-factor integrands, including the mass-dependent
Lorentz projection factors.
In the $q^+=0$ frame, the extractions using the
$(+,\perp)$ and $(+,-)$ current-component combinations
were verified numerically to yield the same $f_-(q^2)$
after integration, without adding separate instantaneous or
zero-mode terms to the mapped on-shell expressions.

The recent study in Ref.~\cite{BTChoi} developed a direct
BT formulation of these pseudoscalar transition form factors.
Starting from meson states constructed within the BT framework
and an on-shell one-body current, it derived the helicity kernels
and internal Lorentz projections without relying on
BS-to-LFQM matching.
In this realization of the BT
construction~\cite{BT53,Poly10,CCKP}, interactions enter
the mass operator and bound-state wave functions, while the
current kernel is built from on-shell constituent spinors.
The fully BT-based LFQM (fBT-LFQM) uses the free constituent
invariant masses consistently in the internal Lorentz projection,
while retaining physical meson masses in external decay
kinematics and in the definition of $f_0$.
For the same wave functions and constituent inputs, the
resulting on-shell form-factor expressions are equivalent
to those obtained through the correspondence in
Ref.~\cite{Choi2021}.
The direct formulation therefore makes explicit the BT
structure already present in that earlier calculation.
It also establishes analytically the equality of the two
independently integrated current-component extractions of
$f_-(q^2)$ for rotationally invariant radial wave functions
with the LF Jacobian and the adopted pseudoscalar spin coupling,
provided the reduced overlaps converge~\cite{BTChoi}.

The present work is a phenomenological follow-up to
Ref.~\cite{BTChoi} that adopts its direct fBT formulation
as the common framework for a systematic comparison
of five potential-model implementations of the QCD-motivated LFQM.
It extends the earlier $B\to D$ analysis of
Ref.~\cite{Choi2021} to the five semileptonic transition
channels considered here.

The parameter sets of these implementations are adopted from
previous spectroscopic analyses~\cite{Choi2009,ZiyueMass,JafarMass,Pandya2024}
and held fixed throughout the decay calculations, without
readjustment to semileptonic data.
We compare the complete implementations to assess the combined
effects of confinement, radial mixing, hyperfine treatment,
and constituent parameters on form-factor normalizations,
recoil dependence, branching fractions, and lepton-flavor ratios.

The timelike overlaps are evaluated directly over the full
physical $q^2$ range of each transition, including zero recoil,
using the continuation developed in Ref.~\cite{BTChoi}.
Independent current-component extractions provide a numerical
consistency check for each implementation and transition.
Predictions are summarized by equal-weight arithmetic means
and minimum--maximum ranges across the five implementations.
Experimental measurements and available lattice results
serve as external comparisons.

Section~\ref{sec:model} summarizes the BT-based wave functions
and the five potential-model implementations.
Section~\ref{sec:formfactors} specifies the form-factor kernels,
direct timelike evaluation, and decay observables.
Section~\ref{sec:results} presents the numerical inputs,
normalization and averaging conventions, and comparative results.
Section~\ref{SUM} summarizes the phenomenological conclusions.
Appendix~\ref{app:bs-bt} contrasts the on-shell fBT overlap
with the triangle-loop calculation in the covariant BS model,
emphasizing their different treatments of the mass-dependent
Lorentz prefactors.
Appendix~\ref{app:auxiliary} presents an auxiliary rational
parametrization of the directly calculated form factors and
quantifies its deviations from the direct-overlap results.

\section{BT-based Light-Front Quark Model}\label{sec:model}
In the BT-based LFQM, each meson is described as a constituent
quark--antiquark ($q\bar q$) bound state~\cite{BTChoi}.
The five potential-model implementations share a common
state construction but differ in their radial wave functions
and constituent parameters.

\subsection{Light-front wave functions}
\label{subsec:lf-wave-functions}

The physical four-momentum $P$ of a meson in LF coordinates is written as
$P=(P^+, P^-, {\bm P}_\perp)$, where $P^{+}=P^0 + P^3$ is the longitudinal LF
momentum, $P^-= P^0-P^3$ is the LF energy, and ${\bm P}_\perp=(P^1, P^2)$ denotes the
transverse components. We adopt the LF metric convention in which the inner
product of two four-vectors is given by
$a\cdot b=\frac{1}{2}(a^+b^- + a^-b^+) - {\bm a}_\perp\cdot {\bm b}_\perp$.
A physical meson state $\ket{P}$ with total momentum $P$, spin $J$, and spin
projection $J_z$ is constructed from its constituent quark and antiquark as
\begin{align}\label{eq:18}
\ket{P}
={}&
\int [\mathrm d^3\bm p_q]\,[\mathrm d^3\bm p_{\bar q}]\,
2(2\pi)^3\,
\delta^{(3)}(\bm P-\bm p_q-\bm p_{\bar q})
\nonumber\\
&\times
\sum_{\lambda_q,\lambda_{\bar q}}
\Psi^{JJ_z}_{\lambda_q\lambda_{\bar q}}(x,\bm k_\perp)\,
\ket{q(p_q,\lambda_q)\bar q(p_{\bar q},\lambda_{\bar q})},
\end{align}
where $\Psi^{JJ_z}_{\lambda_q\lambda_{\bar q}}(x,\bm k_\perp)$
is the LF momentum-space wave function of the meson in the
constituent quark--antiquark basis.
The quantities $p_i=(p_i^+,p_i^-,\bm p_{i\perp})$ and $\lambda_i$
denote the on-shell four-momentum and helicity of constituent
$i=q,\bar q$, respectively.
Boldface momenta denote the three LF momentum components,
$\bm p_i\equiv(p_i^+,\bm p_{i\perp})$ and
$\bm P\equiv(P^+,\bm P_\perp)$.
The integration measure is
$[\mathrm d^3\bm p_i]
=\mathrm dp_i^+\,\mathrm d^2\bm p_{i\perp}/[2(2\pi)^3]$.
The constituent states are normalized as
$
\langle q(p',\lambda')|q(p,\lambda)\rangle
=
2(2\pi)^3\delta_{\lambda'\lambda}
\delta^{(3)}(\bm p'-\bm p)$,
with the same convention for antiquark states.
The three-dimensional delta function in Eq.~\eqref{eq:18}
fixes the total plus and transverse momenta of the constituent
pair to those of the meson.
For each free on-shell basis configuration, we denote the pair
four-momentum by $P_{\rm on}\equiv p_q+p_{\bar q}$.
Its plus and transverse components satisfy
$P_{\rm on}^+=P^+$ and
$\bm P_{{\rm on}\perp}=\bm P_\perp$.
The free invariant mass of the pair is defined by
$M_0^2\equiv P_{\rm on}^2$.
The on-shell conditions then give the free-pair LF energy as
\begin{equation}
\begin{aligned}
P^-_{\rm on}&\equiv p_q^-+p_{\bar q}^-\\
&=\frac{m_q^2+\bm p_{q\perp}^2}{p_q^+}
 +\frac{m_{\bar q}^2+\bm p_{\bar q\perp}^2}{p_{\bar q}^+}
 =\frac{M_0^2+\bm P_\perp^2}{P^+}.
\end{aligned}
\label{eq:MM0}
\end{equation}
The subscript ``on'' distinguishes this free-pair four-momentum
from the physical meson four-momentum $P$.
Equation~\eqref{eq:MM0} expresses LF energy additivity in the
free on-shell constituent basis used for the spinor matrix
elements.
This free-pair energy is distinct from the LF energy eigenvalue
of the interacting meson, which is determined by the BT mass
operator described in Sec.~\ref{subsec:bt-potential-models}.

The LF internal momentum variables are defined by
$x_i=p_i^+/P^+$ and
$\bm k_{i\perp}=\bm p_{i\perp}-x_i\bm P_\perp$,
with $\sum_i x_i=1$ and $\sum_i\bm k_{i\perp}=0$.
For the quark, we choose $x_q\equiv x$ and
$\bm k_{q\perp}\equiv-\bm k_\perp$.
The momentum dependence of the LFWF is thus expressed
entirely in terms of the internal variables $(x,\bm k_\perp)$.
In terms of these variables, the free invariant mass squared is
\begin{equation}\label{eq:M0}
M_0^2 = \frac{\bm k_\perp^2+m_q^2}{x}
      + \frac{\bm k_\perp^2+m_{\bar q}^2}{1-x},
\end{equation}
where $m_{q(\bar q)}$ is the mass of the constituent quark
(antiquark).

We normalize the meson state as
\begin{equation}\label{eq:statenorm}
\Braket{P'|P}=2 (2\pi)^3 P^+\,\delta^{(3)}({\bm P}'-{\bm P}),
\end{equation}
so that the LFWF of a pseudoscalar meson $(J=0,J_z=0)$ satisfies
\begin{equation}\label{eq:Psinorm}
\int^1_0 \mathrm dx\int \frac{\mathrm d^2{\bm k}_\perp}{2(2\pi)^3}
\sum_{\lambda_q,\lambda_{\bar q}}
\Psi^{00\,\dagger}_{\lambda_q\lambda_{\bar q}}(x,{\bm k}_{\perp})
\Psi^{00}_{\lambda_q\lambda_{\bar q}}(x,{\bm k}_{\perp})=1.
\end{equation}

The LF wave function of an $nS$ pseudoscalar meson can be written in the factorized form
\begin{eqnarray}\label{eq:5}
\Psi^{JJ_z}_{\lambda_q\lambda_{\bar q}}(x,{\bm k}_{\perp})
=
\Phi_{nS}(x,{\bm k}_{\perp})\,
\mathcal{R}^{JJ_z}_{\lambda_q\lambda_{\bar q}}(x,{\bm k}_{\perp}),
\end{eqnarray}
where $\Phi_{nS}$ is the radial wave function and
$\mathcal{R}^{JJ_z}_{\lambda_q\lambda_{\bar{q}}}$ is the spin--orbit wave function,
which encodes the coupling between the constituent spins
and the internal orbital motion in the LF helicity basis.
It is obtained through the interaction-independent
Melosh transformation~\cite{Melosh1974},
which relates instant-form spin states to LF helicity states
while preserving the total angular momentum quantum numbers.

For a pseudoscalar ($J^P=0^-$) meson,
the spin--orbit wave function takes the covariant form~\cite{Jaus1990}
\begin{eqnarray}\label{eq8}
\mathcal{R}^{00}_{\lambda_q\lambda_{\bar{q}}} 
&=&
-\frac{\bar{u}_{\lambda_q}(p_q) \gamma_5 v_{\lambda_{\bar{q}}}(p_{\bar{q}})}
{\sqrt{2}\sqrt{M_0^2 - (m_q -m_{\bar{q}})^2}},
\end{eqnarray}
which satisfies the normalization condition
\begin{equation}
\sum_{\lambda_q,\lambda_{\bar q}}
\mathcal{R}^{00\,\dagger}_{\lambda_q\lambda_{\bar q}}
\mathcal{R}^{00}_{\lambda_q\lambda_{\bar q}}=1.
\end{equation}

In the LF helicity basis, Eq.~\eqref{eq8} reduces to the compact matrix form
\begin{equation}\label{Rmatrix}
\mathcal{R}^{00}_{\lambda_q{\lambda_{\bar q}}} =
\frac{1}{\sqrt{2}\sqrt{{\bm k}_\perp^2 + \mathcal{A}^2}}
\begin{pmatrix}
k^L & \mathcal{A} \\
- \mathcal{A} & k^R
\end{pmatrix},
\end{equation}
where $k^{R(L)} = k_x \pm i k_y$ and
$\mathcal{A} = (1 - x) m_q + x m_{\bar q}$.
Rows and columns label the quark and antiquark helicities, respectively,
in the order $(\uparrow,\downarrow)$.\footnote{The diagonal numerators are $(+k^L,+k^R)$ for the present
choice $\bm k_{q\perp}=-\bm k_\perp$ and $(-k^L,-k^R)$ for $\bm k_{q\perp}=+\bm k_\perp$.}
For the ground-state transitions considered here, the radial
wave function is expressed as a linear combination of the two
lowest $S$-wave harmonic-oscillator (HO) basis
functions~\cite{Choi2009,ZiyueMass,JafarMass,Pandya2024}:
\begin{equation}
\Phi_{1S}=\sum_{m=1}^{2}c_{1m}\phi_{mS},
\qquad c_{11}^{2}+c_{12}^{2}=1.
\label{eq:radialtruncation}
\end{equation}
The single-basis implementations correspond to
$(c_{11},c_{12})=(1,0)$, while the mixed-basis implementations
retain both components in a coherent superposition, with
the coefficients listed in Table~\ref{tab:modelsummary}.
In all cases, $\Phi_{1S}$ describes a ground-state meson.
The five potential-model implementations are described in
Sec.~\ref{subsec:bt-potential-models}.
The basis functions are given by
\begin{equation}\label{eq:9}
\begin{aligned}
\phi_{1S}(x,\bm k_\perp)
&=
\frac{4\pi^{3/4}}{\beta^{3/2}}
\sqrt{\frac{\partial k_z}{\partial x}}
\,e^{-\frac{\vec k^{\,2}}{2\beta^2}}, \\
\phi_{2S}(x,\bm k_\perp)
&=
\frac{4\pi^{3/4}\left(2\vec k^{\,2}-3\beta^2\right)}
{\sqrt{6}\beta^{7/2}}
\sqrt{\frac{\partial k_z}{\partial x}}
\,e^{-\frac{\vec k^{\,2}}{2\beta^2}},
\end{aligned}
\end{equation}
where $\vec k=(k_z,\bm k_\perp)$, and $\beta$ denotes
the variational radial-width parameter determined from
meson spectroscopy.
The $z$ component of the internal three-momentum ${\vec k}$ is related to the LF
variables $(x,{\bm k}_\perp)$ through
\begin{equation}
k_z = \left(x - \tfrac{1}{2}\right) M_0
+ \frac{m_{\bar q}^2 - m_q^2}{2 M_0},
\end{equation}
and the change of variables $\{k_z,{\bm k}_\perp\}\to\{x,{\bm k}_\perp\}$ introduces
the Jacobian factor
\begin{equation}
\frac{\partial k_z}{\partial x}
=
\frac{M_0}{4x(1-x)}
\left[
1-\left(\frac{m_q^2-m_{\bar q}^2}{M_0^2}\right)^2
\right].
\end{equation}
The basis functions $\phi_{nS}$ include the square root of the Jacobian,
$\sqrt{\partial k_z/\partial x}$, and satisfy the normalization condition
\begin{eqnarray}\label{eq:10}
 \int_0^1 \mathrm dx \int \frac{\mathrm d^2 {\bm k}_\perp}{2(2\pi)^3}
 \left|\phi_{nS}(x,{\bm k}_\perp)\right|^2 =1.
\end{eqnarray}
The orthonormality of the HO basis, the coefficient
normalization $c_{11}^2+c_{12}^2=1$, and the pointwise
spin normalization
$\sum_{\lambda_q,\lambda_{\bar q}}
|\mathcal R^{00}_{\lambda_q\lambda_{\bar q}}|^2=1$
give Eq.~\eqref{eq:Psinorm} for the full pseudoscalar
wave function.

\subsection{BT mass operator and potential models}
\label{subsec:bt-potential-models}

The state representation and normalized radial basis specified above
are common to the five implementations.
Their binding dynamics are described by the QCD-motivated
effective Hamiltonians employed in previous spectroscopic
analyses within the LFQM~\cite{Choi1997,Choi1999,Choi2009,ZiyueMass,JafarMass,Dhiman2019,Pandya2024}.
These Hamiltonians are defined in the quark--antiquark
center-of-mass frame, where the meson is at rest and its
total energy equals its mass.
They therefore represent the internal invariant mass operator.
We denote this operator by $\hat M$ for each implementation
and write
\begin{equation}
\begin{aligned}
\hat M&=\hat M_0+\hat V_{q\bar q},\\
\hat M_0&=\sqrt{m_q^2+\hat{\vec k}^{\,2}}
     +\sqrt{m_{\bar q}^2+\hat{\vec k}^{\,2}}.
\end{aligned}
\label{eq:spectroscopic-Hamiltonian}
\end{equation}
Here $\hat M_0$ is the free mass operator of the on-shell
quark--antiquark pair, and the corresponding invariant mass
$M_0$ is defined in LF variables in Eq.~\eqref{eq:M0}.
The operator $\hat V_{q\bar q}$ describes the effective
quark--antiquark interaction.
For an exact bound state, $\hat M\ket{\Psi}=M\ket{\Psi}$.
In the spectroscopic analyses underlying the five
implementations, the radial wave functions were constructed
from the finite HO basis specified above and used as trial
functions in applying the variational principle.
The radial width $\beta$ was taken as the variational
parameter.

In the BT construction, the kinematic LF generators retain their
free-particle form.
The interaction is constrained, including by commutation with
the noninteracting internal spin, so that the resulting
generators satisfy the Poincar\'e algebra~\cite{BT53,Poly10,CCKP}.
At fixed total $P^+$ and $\bm P_\perp$, the LF energy
generator $\hat P^-$ acts on a meson mass eigenstate as
\begin{equation}
\hat P^-\ket{P}
=\frac{M^2+\bm P_\perp^2}{P^+}\ket{P}.
\label{eq:meson-LF-energy}
\end{equation}
The eigenvalue $P^-$ is the physical meson LF
energy~\cite{ZiyueMass}.
The spectroscopic analyses use the internal mass operator
$\hat M$ to determine the meson masses and radial wave
functions independently of the overall meson momentum.
For exact states, the $\hat M$ and $\hat P^-$ eigenvalue
problems are equivalent.
By contrast, $P^-_{\rm on}$ in Eq.~\eqref{eq:MM0} is the
free-pair LF energy associated with each on-shell
constituent basis configuration.

This distinction is essential to the one-body current calculation.
The binding dynamics enter through the meson wave functions,
while the current kernel is evaluated between free on-shell
constituent states.
In the fBT realization adopted here, the mass-dependent Lorentz
prefactors in the internal form-factor projection must
therefore be evaluated with the same free-pair kinematics as
the on-shell spinor matrix elements~\cite{Arifi2022,BTChoi}.
For a transition, this means using $M_0$ and $M'_0$
in place of $M$ and $M'$, respectively, in the internal
prefactors at each integration point.

This prescription does not replace the interacting mass
operator $\hat M$ by $\hat M_0$ or remove binding from
the meson state.
Physical meson masses are retained in the external hadron
kinematics and in the definition of $f_0$.
The Poincar\'e consistency of the BT state construction does
not by itself establish full Lorentz covariance of the
approximate one-body current~\cite{BTChoi}.

The five implementations differ in their
confinement potentials, radial basis composition, and
treatment of the hyperfine interaction.
The \mbox{Lin-1S-P} and \mbox{HO-1S-P} implementations use
linear and HO confinement, respectively,
with the single HO radial basis function $\phi_{1S}$
and perturbative hyperfine corrections~\cite{Choi2009}.
The \mbox{Lin-Mix-NP} implementation combines linear
confinement with a coherent superposition of $\phi_{1S}$
and $\phi_{2S}$ and includes a Gaussian-smeared hyperfine
interaction in the variational
optimization~\cite{ZiyueMass}.
The \mbox{Lin-Mix-P} and \mbox{Log-Mix-P} implementations
also employ mixed radial wave functions, with linear
and logarithmic confinement, respectively, and perturbative
hyperfine corrections~\cite{JafarMass,Pandya2024}.
In these labels, ``1S'' and ``Mix'' denote the single-basis
and mixed-basis radial wave functions, respectively.
P and NP denote perturbative and nonperturbative treatments
of the hyperfine interaction, respectively.

Let $\vec r$ be the relative coordinate conjugate to $\vec k$, with $r=|\vec r|$.
We denote the coordinate-space expressions of the potential
operators $\hat V_i$ by $V_i(r)$.
The common decomposition is
\begin{equation}
V_{q\bar q}(r)
=
V_{\mathrm{conf}}(r)
+V_{\mathrm{Coul}}(r)
+V_{\mathrm{hyp}}(r).
\label{eq:potential-decomposition}
\end{equation}
The Coulomb term has the same functional form in all five
implementations,
\begin{equation}
V_{\mathrm{Coul}}(r)=-\frac{4\alpha_s}{3r},
\label{eq:VCoul}
\end{equation}
while the three confinement forms are expressed in a common
notation as~\cite{Choi2009,ZiyueMass,JafarMass,Pandya2024}
\begin{equation}
V_{\mathrm{conf}}(r)=a+
\begin{cases}
br, & \text{Lin-1S-P, Lin-Mix-NP, Lin-Mix-P},\\[2pt]
br^2, & \text{HO-1S-P},\\[2pt]
c\ln(r/r_0), & \text{Log-Mix-P}.
\end{cases}
\label{eq:confinement-forms}
\end{equation}
Here $a$ is an additive constant, $b$ and $c$ are confinement
strengths, $r_0$ is a reference length, and $\alpha_s$ is an
effective coupling fixed in each spectroscopic analysis.
The same symbols are used for model-specific parameters
and do not imply common numerical values across implementations.
In particular, $b$ has dimensions of energy squared for
linear confinement and energy cubed for HO
confinement in natural units.

For $S$-wave mesons, the spin--spin part of the effective
one-gluon-exchange interaction has the contact form
\begin{equation}
V_{\mathrm{hyp}}^{\mathrm{ct}}(r)
=
\frac{2\,\bm S_q\!\cdot\!\bm S_{\bar q}}{3m_qm_{\bar q}}
\nabla_r^2 V_{\mathrm{Coul}}(r)
=
\frac{32\pi\alpha_s}{9m_qm_{\bar q}}
\bm S_q\!\cdot\!\bm S_{\bar q}\,
\delta^{(3)}(\vec r),
\label{eq:hyperfine-contact}
\end{equation}
where
$\nabla_r^2 V_{\mathrm{Coul}}
=(16\pi\alpha_s/3)\delta^{(3)}(\vec r)$.
The spin expectation value is
$\langle\bm S_q\cdot\bm S_{\bar q}\rangle=-3/4$
for pseudoscalar mesons and $1/4$ for vector mesons.
The four implementations labeled P treat the unsmeared
contact hyperfine interaction perturbatively, with their
respective confinement forms, radial trial functions, and
spectroscopy-based parameters~\cite{Choi2009,JafarMass,Pandya2024}.
Their radial widths were determined from the central part
of the mass operator,
$\hat M_{\mathrm{cent}}
=\hat M_0+\hat V_{\mathrm{conf}}+\hat V_{\mathrm{Coul}}$,
through
\begin{equation}
\frac{\partial}{\partial\beta}
\langle\hat M_{\mathrm{cent}}\rangle_\beta=0
\qquad (\mathrm{P}),
\label{eq:variation-P}
\end{equation}
and $\langle\hat V_{\mathrm{hyp}}^{\mathrm{ct}}\rangle_\beta$
was then added as a first-order mass
correction~\cite{Choi2009,JafarMass,Pandya2024}.
Here the brackets denote an expectation value in the
normalized trial state of specified spin, and the derivative
is taken at fixed constituent masses, potential parameters,
and mixing amplitudes.
In this prescription, the optimized radial width is common
to the pseudoscalar and vector partners of a given flavor
content and radial state.

The \mbox{Lin-Mix-NP} implementation of Ref.~\cite{ZiyueMass}
retains linear confinement and the same unsmeared Coulomb
form in Eq.~\eqref{eq:VCoul}.
In this implementation, only the hyperfine contact distribution
is smeared before the variational minimization:
\begin{equation}
\begin{aligned}
\delta^{(3)}(\vec r)
&\longrightarrow
g_{\sigma_{q\bar q}}(r)
=
\frac{\sigma_{q\bar q}^{3}}{\pi^{3/2}}
e^{-\sigma_{q\bar q}^{2}r^2},\\[2pt]
V_{\mathrm{hyp}}^{\mathrm{sm}}(r)
&=
\frac{32\pi\alpha_s}{9m_qm_{\bar q}}
\bm S_q\!\cdot\!\bm S_{\bar q}\,
g_{\sigma_{q\bar q}}(r).
\end{aligned}
\label{eq:hyperfine-smeared}
\end{equation}
The Gaussian is normalized to unity in three-dimensional
coordinate space.
The flavor-dependent smearing scale is denoted by
$\sigma_{q\bar q}=\lambda_{q\bar q}\sigma$.
The spectroscopic analysis of Ref.~\cite{ZiyueMass} uses
$\sigma=0.423~\mathrm{GeV}$,
$\lambda_{c\bar c}=2$, $\lambda_{b\bar c}=2.3$, and
$\lambda_{b\bar b}=3$, with $\lambda_{q\bar q}=1$
for the remaining flavor sectors.
The smeared hyperfine interaction is included in the
variational mass operator, giving the stationarity condition
\begin{equation}
\frac{\partial}{\partial\beta}
\left\langle
\hat M_{\mathrm{cent}}+\hat V_{\mathrm{hyp}}^{\mathrm{sm}}
\right\rangle_\beta=0
\qquad (\mathrm{NP}).
\label{eq:variation-NP}
\end{equation}
The smearing scale is also held fixed when taking this
derivative with respect to $\beta$.
Here, ``nonperturbative'' (NP) refers specifically to the
inclusion of the smeared hyperfine interaction in the
variational optimization within the adopted trial space.
The spin dependence of this interaction yields different
optimal radial widths for pseudoscalar and vector mesons
of the same flavor content~\cite{ZiyueMass}.
Only the pseudoscalar widths are used in the present
decay calculation.

\begin{table*}[!htb]
\centering
\small
\caption{Characteristics of the five LFQM implementations.
The labels identify the confinement form, the ground-state
HO-basis composition, and the hyperfine prescription.
The radial amplitudes use the phase convention of Eq.~\eqref{eq:9}.
The explicit potential terms and the P/NP variational
prescriptions are given in
Eqs.~\eqref{eq:potential-decomposition}--\eqref{eq:variation-NP}.}
\label{tab:modelsummary}
\setlength{\tabcolsep}{4pt}
\renewcommand{\arraystretch}{1.45}
\begin{tabular*}{\textwidth}{@{\extracolsep{\fill}}lclll@{}}
\hline\hline
Model & $V_{\mathrm{conf}}-a$ & $\Phi_{1S}$
& $V_{\mathrm{hyp}}$ & Treatment of $V_{\mathrm{hyp}}$ \\
\hline
\mbox{Lin-1S-P}~\cite{Choi2009}
& $br$
& $\phi_{1S}$
& Contact
& Perturbative (P) \\

\mbox{HO-1S-P}~\cite{Choi2009}
& $br^2$
& $\phi_{1S}$
& Contact
& Perturbative (P) \\

\mbox{Lin-Mix-NP}~\cite{ZiyueMass}
& $br$
& $\sqrt{0.7}\,\phi_{1S}+\sqrt{0.3}\,\phi_{2S}$
& Gaussian-smeared
& \shortstack[l]{Included in\\variation (NP)} \\

\mbox{Lin-Mix-P}~\cite{JafarMass}
& $br$
& $\cos12^\circ\,\phi_{1S}+\sin12^\circ\,\phi_{2S}$
& Contact
& Perturbative (P) \\

\mbox{Log-Mix-P}~\cite{Pandya2024}
& $c\ln(r/r_0)$
& $\cos18^\circ\,\phi_{1S}+\sin18^\circ\,\phi_{2S}$
& Contact
& Perturbative (P) \\
\hline\hline
\end{tabular*}
\end{table*}

\begin{table*}[!htb]
\centering
\small
\caption{Constituent quark masses and pseudoscalar radial
widths $\beta_{q\bar q}$ for the five implementations
characterized in Table~\ref{tab:modelsummary}.
All values are in GeV, and $q=u,d$.}
\label{input}
\setlength{\tabcolsep}{9pt}
\renewcommand{\arraystretch}{1.45}
\begin{tabular}{lcccccccccc}
\hline\hline
Model & $m_q$ & $m_s$ & $m_c$ & $m_b$
& $\beta_{qc}$ & $\beta_{sc}$ & $\beta_{cc}$
& $\beta_{qb}$ & $\beta_{sb}$ & $\beta_{cb}$ \\
\hline
\mbox{Log-Mix-P}~\cite{Pandya2024}
& 0.22 & 0.45 & 1.68 & 5.10
& 0.4435 & 0.4763 & 0.6193
& 0.5407 & 0.5820 & 0.7882 \\

\mbox{Lin-Mix-P}~\cite{JafarMass}
& 0.22 & 0.45 & 1.68 & 5.10
& 0.4240 & 0.4550 & 0.5920
& 0.4950 & 0.5380 & 0.7670 \\

\mbox{Lin-Mix-NP}~\cite{ZiyueMass}
& 0.205 & 0.38 & 1.75 & 5.15
& 0.3801 & 0.3859 & 0.5270
& 0.4226 & 0.4412 & 0.6646 \\

\mbox{HO-1S-P}~\cite{Choi2009}
& 0.25 & 0.48 & 1.80 & 5.20
& 0.4216 & 0.4686 & 0.6998
& 0.4960 & 0.5740 & 1.0350 \\

\mbox{Lin-1S-P}~\cite{Choi2009}
& 0.22 & 0.45 & 1.80 & 5.20
& 0.4679 & 0.5016 & 0.6509
& 0.5266 & 0.5712 & 0.8068 \\
\hline\hline
\end{tabular}
\end{table*}

In \mbox{Lin-Mix-NP}, the two-basis expansion was used
as a variational ansatz for the ground-state radial
wave function~\cite{ZiyueMass}.
In \mbox{Lin-Mix-P} and \mbox{Log-Mix-P}, the original
spectroscopic analyses instead employed orthogonal rotations
of $(\phi_{1S},\phi_{2S})$ to describe the ground and first
radially excited states simultaneously.
The mass-gap relation and decay-constant hierarchy were
used to constrain the mixing~\cite{JafarMass,Pandya2024}.
Only the resulting ground-state radial wave functions
are used in the present decay calculation.

Tables~\ref{tab:modelsummary} and~\ref{input} summarize the
model characteristics and numerical inputs, respectively,
for the present semileptonic overlap calculations.
The potential forms and variational conditions given above
describe the spectroscopic origin of these inputs.
The effects of the model interactions enter the decay
calculation through the radial LFWFs and constituent parameters.
All five implementations use the same functional form of the
on-shell weak-current kernel and the same internal fBT
projection prescription, evaluated with their respective inputs.

\section{Form Factors for Semileptonic Decays}\label{sec:formfactors}

\subsection{Physical kinematics and the fBT prescription}

We consider a pseudoscalar transition
$P(q_1\bar q)\to P'(q_2\bar q)$, with initial and final
meson momenta $P$ and $P'$ and masses $M$ and $M'$,
respectively.
The physical weak vector-current matrix element is
parametrized as
\begin{equation}
\langle J^\mu\rangle
\equiv\langle P'|\bar q_2\gamma^\mu q_1|P\rangle
=f_+(q^2)\,\mathcal P^\mu+f_-(q^2)\,q^\mu,
\label{eq:Jdef}
\end{equation}
where $\mathcal P=P+P'$ and $q=P-P'$.
The physical decay interval is
$m_\ell^2\le q^2\le(M-M')^2$, where $m_\ell$ is the
charged-lepton mass.
The scalar form factor is defined with the physical
meson masses,
\begin{equation}
f_0(q^2)
=f_+(q^2)+\frac{q^2}{M^2-M'^2}f_-(q^2),
\label{eq:f0def}
\end{equation}
so that $f_0(0)=f_+(0)$.
Equivalently,
\begin{equation}
\langle J^\mu\rangle
=f_+(q^2)
\left(
\mathcal P^\mu-\frac{\mathcal P\cdot q}{q^2}q^\mu
\right)
+f_0(q^2)\frac{\mathcal P\cdot q}{q^2}q^\mu,
\label{eq:transdecomp}
\end{equation}
with $\mathcal P\cdot q=M^2-M'^2$.
For $q^2\ne0$, the two terms are transverse and
longitudinal with respect to $q^\mu$, respectively.
Contracting with $q_\mu$ gives
$q_\mu\langle J^\mu\rangle=(M^2-M'^2)f_0(q^2)$.
The condition $f_0(0)=f_+(0)$ removes the apparent
kinematic singularity at $q^2=0$.

We work in the Drell--Yan--West (DYW) frame,
defined by $q^+=0$, and choose
$\bm P_\perp=\bm 0_\perp$.
The physical external momenta are
\begin{align}
P&=\left(P^+,\frac{M^2}{P^+},\bm 0_\perp\right),
\nonumber\\
P'&=\left(
P^+,\frac{M'^2+\bm q_\perp^2}{P^+},-\bm q_\perp
\right),
\nonumber\\
q&=\left(
0,\frac{M^2-M'^2-\bm q_\perp^2}{P^+},\bm q_\perp
\right).
\label{eq:externalmomenta}
\end{align}
This choice gives $\mathcal P^+=2P^+$ and
$q^2=-\bm q_\perp^2\le0$.
The form factors in the physical timelike region are
obtained by direct analytic continuation of the overlap,
as described below.

We adopt the fBT prescription of Ref.~\cite{BTChoi}.
As discussed in Sec.~\ref{sec:model}, the one-body current
kernel is constructed from free on-shell constituent spinors.
The associated total free-pair momenta $P_{\rm on}$ and
$P'_{\rm on}$ have invariant masses
$M_0(x,\bm k_\perp)$ and $M'_0(x,\bm k'_\perp)$,
respectively.
The mass-dependent Lorentz prefactors are evaluated with
these same free-pair kinematics at each integration point.
Defining $q_{\rm on}=P_{\rm on}-P'_{\rm on}$ and
$\mathcal P_{\rm on}=P_{\rm on}+P'_{\rm on}$, the
minus-current projection uses
\begin{equation}
\begin{aligned}
q^-_{\rm on}
&=P^-_{\rm on}-P_{\rm on}^{\prime-}
 =\frac{M_0^2-M_0^{\prime2}-\bm q_\perp^2}{P^+},\\
\mathcal P^-_{\rm on}
&=P^-_{\rm on}+P_{\rm on}^{\prime-}
 =\frac{M_0^2+M_0^{\prime2}+\bm q_\perp^2}{P^+}.
\end{aligned}
\label{eq:onprefactors}
\end{equation}
With the spectator momentum unchanged,
$q_{\rm on}=p_1-p_2$ is also the momentum transfer
between the initial and final active quarks, whose
on-shell momenta are denoted by $p_1$ and $p_2$.
Its plus and transverse components coincide with those
of the physical $q$, while its minus component is fixed
by the constituent on-shell energies.
Consequently, $q_{\rm on}^2=q^2$ in the $q^+=0$ frame,
although $q^-_{\rm on}$ generally differs from $q^-$.

The replacements $M\to M_0$ and $M'\to M'_0$ thus
match the internal Lorentz projection to the on-shell
current kernel.
They do not identify the physical meson energy with
the free-pair energy.
The mass-dependent projection is performed inside the
overlap integral, and the resulting projected kernels
are integrated with the LFWFs.

The plus component fixes $f_+$, while combining it with
either the transverse or the minus current yields two
independent extractions of $f_-$.
We denote these extractions by $f_-^{(\perp)}$ and $f_-^{(-)}$,
respectively.\footnote{A detailed comparison with the partially BT-based (pBT)
and covariant BS formulations is given in Ref.~\cite{BTChoi}.}

After integration, the extracted form factors are the invariant
coefficients of the physical
hadron-level matrix element in Eq.~\eqref{eq:Jdef}.
The external momenta satisfy $P^2=M^2$ and $P'^2=M'^2$.
The physical meson masses $M$ and $M'$ are also used in
the scalar combination in Eq.~\eqref{eq:f0def}, the decay
endpoint, and the phase-space factors.

\subsection{Form factors in the fBT-LFQM}

Using the internal fBT projection specified above, we
obtain the LFWF overlap expressions for $f_+(q^2)$ and
$f_-(q^2)$~\cite{BTChoi}.
For the transition $P(q_1\bar q)\to P'(q_2\bar q)$
in the DYW frame, the plus and transverse momenta
of the active quarks and spectator antiquark are assigned as
\begin{alignat}{2}
p_1^+ &= xP^+,
&\qquad p_2^+ &= xP^+, \nonumber\\
p_{\bar q}^+ &= (1-x)P^+,
&\qquad p_{\bar q}^{\prime+} &= (1-x)P^+, \nonumber\\
\bm p_{1\perp} &= x\bm P_\perp-\bm k_\perp,
&\qquad \bm p_{2\perp}
&=x\bm P'_\perp-\bm k'_\perp, \nonumber\\
\bm p_{\bar q\perp} &= (1-x)\bm P_\perp+\bm k_\perp,
&\qquad \bm p'_{\bar q\perp}
&=(1-x)\bm P'_\perp+\bm k'_\perp.
\label{eq:LFmomenta}
\end{alignat}
The spectator approximation leaves the antiquark momentum
unchanged,
$p_{\bar q}^{\prime+}=p_{\bar q}^+$ and
$\bm p'_{\bar q\perp}=\bm p_{\bar q\perp}$.
Its minus component is also unchanged because the spectator
remains on shell with the same mass.
With $q=P-P'$, these relations imply
$\bm k'_\perp=\bm k_\perp+(1-x)\bm q_\perp$.

With these momentum assignments, the on-shell one-body
current overlap takes the form
\begin{equation}
\langle J^\mu\rangle_{\rm BT}
=
\int [\mathrm d^3\bm p_1]\,
\Phi'_{1S}(x,\bm k'_\perp)
\Phi_{1S}(x,\bm k_\perp)\,
\mathcal H^\mu,
\label{eq:WeakCurrentConvolution}
\end{equation}
where
$[\mathrm d^3\bm p_1]
=P^+\,\mathrm dx\,\mathrm d^2\bm k_\perp/[2(2\pi)^3]$.
The helicity kernel is defined as
\begin{align}
\mathcal H^\mu
&=
\sum_{\lambda_1,\lambda_2,\bar\lambda}
\mathcal R^{00\dagger}_{\lambda_2\bar\lambda}
(x,\bm k'_\perp)
\left[
\frac{\bar u_{\lambda_2}(p_2)}{\sqrt{p_2^+}}
\gamma^\mu
\frac{u_{\lambda_1}(p_1)}{\sqrt{p_1^+}}
\right]
\mathcal R^{00}_{\lambda_1\bar\lambda}
(x,\bm k_\perp)
\nonumber\\
&\equiv
\sum_{\lambda_1,\lambda_2,\bar\lambda}
\mathcal H^\mu_{\lambda_1\bar\lambda
\to\lambda_2\bar\lambda},
\label{eq:helicity-kernel}
\end{align}
where $\lambda_1$ and $\lambda_2$ denote the helicities
of the initial and final active quarks, respectively,
and $\bar\lambda\equiv\lambda_{\bar q}$ is the
unchanged spectator helicity.
At real momenta, the dagger on a final-state helicity coefficient
denotes complex conjugation of that coefficient.
The bracketed factor is the on-shell vector-current
matrix element between LF helicity spinors.
Its explicit expressions for the plus, transverse, and
minus current components are summarized in
Table~\ref{tab:dirac}.
When the integration measure and helicity kernels are
combined with the corresponding Lorentz projection
factors, their $P^+$ dependence cancels, leaving the
extracted form factors independent of $P^+$.

\begin{table*}[t]
\caption{Dirac matrix elements for the helicity spinors~\cite{LB80}.
Upper and lower signs correspond to the upper and lower transitions
listed in each column, respectively. We use $\epsilon^{xy}=+1$.
Here $p^{L(R)} q^{R(L)} =
\bm{p}_\perp \!\cdot\! \bm{q}_\perp
\pm i(\bm{p}_\perp \!\times\! \bm{q}_\perp)_z$, with $(\bm{p}_\perp \!\times\! \bm{q}_\perp)_z = p_x q_y - p_y q_x$. }
\centering
\renewcommand{\arraystretch}{1.5}
\setlength{\tabcolsep}{15pt}\label{tab:dirac} 
\begin{tabular}{ccc} \hline\hline
  & \multicolumn{2}{c}{Helicity ($\lambda_1\to\lambda_2$)} \\
 Matrix elements    & $\uparrow~\to~\uparrow$ & $\uparrow~\to~\downarrow$ \\
         & $\downarrow~\to~\downarrow$ & $\downarrow~\to~\uparrow$ \\
\hline
$\frac{{\bar u}_{\lambda_2}(p_2)}{\sqrt{p_2^+}}~\gamma^+~\frac{u_{\lambda_1}(p_1)}{\sqrt{p_1^+}}$ & 2 & 0  \\
$\frac{{\bar u}_{\lambda_2}(p_2)}{\sqrt{p_2^+}}~\gamma^-~ \frac{u_{\lambda_1}(p_1)}{\sqrt{p_1^+}}$  & $\frac{2}{{p^{+}_2}{p^+_1}}\left[\bm{p}_{2\perp}\cdot \bm{p}_{1\perp} \pm i (\bm{p}_{2\perp}\times \bm{p}_{1\perp})_z + m_1 m_2 \right]$ 
& $\mp\frac{2}{p^{+}_2 p^+_1}[m_1 (p^x_2 \pm i p^y_2) - m_2 (p^x_1 \pm i p^y_1)]$   \\
$\frac{{\bar u}_{\lambda_2}(p_2)}{\sqrt{p_2^+}}~\gamma^i_{\perp}~ \frac{u_{\lambda_1}(p_1)}{\sqrt{p_1^+}}$   
& $\frac{ \bm{p}^i_{2\perp} \mp i\epsilon^{ij}\bm{p}^j_{2\perp}}{p^{+}_2} + \frac{ \bm{p}^i_{1\perp} \pm i\epsilon^{ij}\bm{p}^j_{1\perp}}{p^+_1}$
&   $\mp\left(\frac{m_1 p^{+}_2 - m_2 p^+_1}{{p^{+}_2}{p^+_1}}\right)(\delta^{ix}\pm i\delta^{iy})$ \\
\hline\hline
\end{tabular}
\end{table*}

\begin{figure*}[t]
\centering
\includegraphics[width=0.85\textwidth]{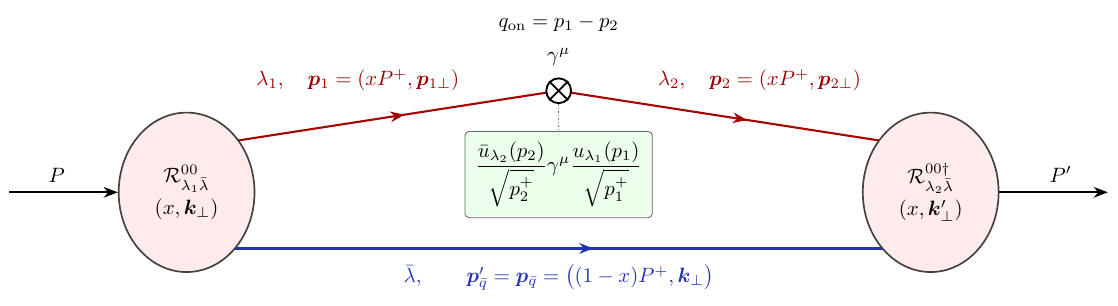}
\caption{Schematic representation of a fixed-helicity
contribution to the on-shell kernel in
Eq.~\eqref{eq:helicity-kernel}, shown in the DYW frame
with $\bm P_\perp=\bm 0_\perp$.
The symbol $\otimes$ marks the one-body vector-current
insertion $\gamma^\mu$.
The label $q_{\rm on}=p_1-p_2$ denotes the momentum
transfer between the on-shell active quarks.}
\label{fig:helicity-kernel}
\end{figure*}

Figure~\ref{fig:helicity-kernel} illustrates an individual
helicity contribution
$\mathcal H^\mu_{\lambda_1\bar\lambda
\to\lambda_2\bar\lambda}$.
The left and right blobs represent the initial and
conjugate final spin--orbit factors, respectively.
The current insertion connects the active-quark helicities
$\lambda_1$ and $\lambda_2$, while the spectator antiquark
retains its momentum and helicity.
In the fBT realization adopted here, the minus-current
projection uses $q^-_{\rm on}$ and $\mathcal P^-_{\rm on}$
in Eq.~\eqref{eq:onprefactors} at each integration point.
This keeps the mass-dependent Lorentz prefactors consistent
with the free-pair kinematics of the on-shell helicity
kernel~\cite{Arifi2022,BTChoi}.
Appendix~\ref{app:bs-bt} contrasts this on-shell prescription
with the covariant BS triangle-loop calculation, where
the constituent propagators are generally off shell and
the Lorentz prefactors retain the physical meson masses.

We separate these contributions according to whether the
active-quark helicity is preserved or flipped at the
current insertion.
The helicity-nonflip (NF) and helicity-flip (HF) terms
satisfy
$\mathcal H^\mu
=\mathcal H^\mu_{\rm NF}+\mathcal H^\mu_{\rm HF}$,
with
\begin{equation}
\mathcal H^\mu_{\rm NF}
=
\sum_{\bar\lambda}
\left(
\mathcal H^\mu_{\uparrow\bar\lambda
\to\uparrow\bar\lambda}
+
\mathcal H^\mu_{\downarrow\bar\lambda
\to\downarrow\bar\lambda}
\right),
\end{equation}
and
\begin{equation}
\mathcal H^\mu_{\rm HF}
=
\sum_{\bar\lambda}
\left(
\mathcal H^\mu_{\uparrow\bar\lambda
\to\downarrow\bar\lambda}
+
\mathcal H^\mu_{\downarrow\bar\lambda
\to\uparrow\bar\lambda}
\right).
\end{equation}
Here $\uparrow$ and $\downarrow$ denote LF helicities
$+1/2$ and $-1/2$, respectively.
The plus current conserves the active-quark LF helicity,
so that $\mathcal H^+_{\rm HF}=0$.
The transverse and minus components generally receive
contributions from both helicity sectors.

Evaluating the helicity sums gives the following explicit
NF and HF kernels~\cite{BTChoi}.
For the transverse current, we display only the contraction
with $\bm q_\perp$, which enters $f_-^{(\perp)}$.
\begin{widetext}
\begin{align}
\mathcal H^+_{\rm NF}
&=
\frac{2\left(\bm k_\perp\cdot\bm k'_\perp
+\mathcal A_1\mathcal A_2\right)}
{\sqrt{\mathcal A_1^2+\bm k_\perp^2}
 \sqrt{\mathcal A_2^2+\bm k_\perp^{\prime\,2}}},
\nonumber\\
\bm{\mathcal H}^{\perp}_{\rm NF}\!\cdot\!\bm q_\perp
&=
-\frac{
\left(\bm k_\perp\cdot\bm k'_\perp
+\mathcal A_1\mathcal A_2\right)
\left(2\bm k_\perp\!\cdot\!\bm q_\perp+\bm q_\perp^2\right)}
{xP^+
 \sqrt{\mathcal A_1^2+\bm k_\perp^2}
 \sqrt{\mathcal A_2^2+\bm k_\perp^{\prime\,2}}},
\nonumber\\
\mathcal H^-_{\rm NF}
&=
\frac{2}{
x^2(P^+)^2
\sqrt{\mathcal A_1^2+\bm k_\perp^2}
\sqrt{\mathcal A_2^2+\bm k_\perp^{\prime\,2}}
}
\bigg[
\left(\bm k_\perp\cdot\bm k'_\perp
+\mathcal A_1\mathcal A_2\right)
\left(\bm k_\perp^2+\bm k_\perp\cdot\bm q_\perp
+m_1m_2\right)
\nonumber\\
&\hspace{4.2cm}
+(1-x)\left(\bm k_\perp\times\bm q_\perp\right)^2
\bigg],
\end{align}
and
\begin{align}
\bm{\mathcal H}^{\perp}_{\rm HF}\!\cdot\!\bm q_\perp
&=
\frac{
\left(m_1p_2^+-m_2p_1^+\right)
\left(\mathcal A_2\bm k_\perp\!\cdot\!\bm q_\perp
-\mathcal A_1\bm k'_\perp\!\cdot\!\bm q_\perp\right)}
{x^2(P^+)^2
 \sqrt{\mathcal A_1^2+\bm k_\perp^2}
 \sqrt{\mathcal A_2^2+\bm k_\perp^{\prime\,2}}},
\nonumber\\
\mathcal H^-_{\rm HF}
&=
\frac{1}{
x^2(P^+)^2
\sqrt{\mathcal A_1^2+\bm k_\perp^2}
\sqrt{\mathcal A_2^2+\bm k_\perp^{\prime\,2}}
}
\bigg[
2(m_2-m_1)(\mathcal A_2-\mathcal A_1)\bm k_\perp^2
+2m_1\mathcal A_1(1-x)\bm q_\perp^2
\nonumber\\
&\hspace{4.2cm}
-2(\bm k_\perp\cdot\bm q_\perp)
\left\{
m_1(\mathcal A_2-\mathcal A_1)
+(m_2-m_1)\mathcal A_1(1-x)
\right\}
\bigg],
\end{align}
\end{widetext}
where $\mathcal A_i=(1-x)m_i+x m_{\bar q}$ for $i=1,2$.
The functional form of these on-shell helicity kernels
is common to all five potential-model implementations.
Each implementation supplies its own constituent masses
and radial wave functions for the overlap calculation.

Following the kernel notation of Ref.~\cite{BTChoi},
we write the form factor $f_+$ and the two extractions
of $f_-$ in the fBT-LFQM as
\begin{align}
f_+(q^2)
&=\int[\mathrm d^3\bm p_1]\,
\Phi'_{1S}(x,\bm k'_\perp)\Phi_{1S}(x,\bm k_\perp)\,
\mathscr K_+,
\label{eq:fplus}\\
f_-^{(\perp)}(q^2)
&=\int[\mathrm d^3\bm p_1]\,
\Phi'_{1S}(x,\bm k'_\perp)\Phi_{1S}(x,\bm k_\perp)\,
\mathscr K_\perp,
\label{eq:fperp}\\
f_-^{(-)}(q^2)
&=\int[\mathrm d^3\bm p_1]\,
\Phi'_{1S}(x,\bm k'_\perp)\Phi_{1S}(x,\bm k_\perp)\,
\mathscr K_-^{\rm on}.
\label{eq:fminus}
\end{align}
The projected current kernels are
\begin{equation}
\begin{aligned}
\mathscr K_+
&=\frac{\mathcal H^+}{\mathcal P^+},
&\qquad
\mathscr K_\perp
&=\mathscr K_+
+\frac{\bm{\mathcal H}^{\perp}\!\cdot\!\bm q_\perp}
{\bm q_\perp^2},\\[4pt]
\mathscr K_-^{\rm on}
&=\frac{\mathcal H^-
-\mathcal P^-_{\rm on}\mathscr K_+}{q^-_{\rm on}}.
\end{aligned}
\label{eq:projectedkernels}
\end{equation}
All kernels depend on the internal variables and the
momentum transfer.
The integration measure retains the factor $P^+$ specified
after Eq.~\eqref{eq:WeakCurrentConvolution}, while
$\mathcal P^+=2P^+$.
No additional factor of $P^+$ is absorbed into
$\mathscr K_+$, $\mathscr K_\perp$, or
$\mathscr K_-^{\rm on}$.
Both $q^-_{\rm on}$ and $\mathcal P^-_{\rm on}$ in
Eq.~\eqref{eq:projectedkernels} are evaluated at each
integration point according to Eq.~\eqref{eq:onprefactors}.

For rotationally invariant radial wave functions with the
LF Jacobian and the adopted pseudoscalar spin coupling,
Ref.~\cite{BTChoi} established the integrated identity
\begin{equation}
f_-^{(\perp)}(q^2)
=f_-^{(-)}(q^2)
=f_-(q^2),
\label{eq:fBTcomponents}
\end{equation}
provided the overlaps converge.
This identity does not require pointwise equality of
$\mathscr K_\perp$ and $\mathscr K_-^{\rm on}$.
The finite HO mixtures introduced in Sec.~\ref{sec:model}
belong to the radial class covered by this result.
We test Eq.~\eqref{eq:fBTcomponents} by independently
integrating the two projected kernels for every
implementation and transition.
This provides a numerical consistency check within
the adopted one-body-current approximation.

The form factors in the physical timelike interval,
$0<q^2\le q^2_{\max}$, are evaluated by direct analytic
continuation of the spacelike overlaps.
We set $\bm q_\perp=(i\sqrt{q^2},0)$ and use the
mass-weighted contour
$\bm k_\perp=\bm\ell_\perp-\zeta(1-x)\bm q_\perp$ and 
$\bm k'_\perp =\bm\ell_\perp+(1-\zeta)(1-x)\bm q_\perp$,
where $\bm\ell_\perp\in\mathbb R^2$ and
$\zeta=m_1/(m_1+m_2)$.
This procedure covers the full physical $q^2$ range of
every model--channel combination, including zero recoil,
without extrapolation or a form-factor fit.
For spacelike transfer, the minus-current overlap is
integrated over $x$ first at fixed, unshifted real
$\bm k_\perp$, so that the pole residue at
$q^-_{\rm on}=0$ cancels before the transverse integration.
At $q^2=0$, the transverse extraction is evaluated as
the finite limit of the complete overlap after angular
integration.
The detailed derivation and regularity analysis are given
in Ref.~\cite{BTChoi}.

\subsection{Differential decay width}

For the semileptonic decay $P\to P'\ell\nu_\ell$, the
differential width with a nonzero charged-lepton mass
$m_\ell$ is expressed in terms of helicity amplitudes
as~\cite{Korner1989,Yao2019}
\begin{equation}
\frac{d\Gamma_\ell}{dq^2}
=
\frac{8|\bm p|\mathcal N(q^2)}{3}
\left[
\left(1+\frac{m_\ell^2}{2q^2}\right)|H_0(q^2)|^2
+\frac{3m_\ell^2}{2q^2}|H_t(q^2)|^2
\right],
\label{eq:helicitywidth}
\end{equation}
where $|\bm p|$ is the magnitude of the daughter-meson
three-momentum in the parent-meson rest frame,
\begin{equation}
|\bm p|
=
\frac{1}{2M}
\sqrt{(M^2+M'^2-q^2)^2-4M^2M'^2}.
\end{equation}
The normalization factor is
\begin{equation}
\mathcal N(q^2)
=
\frac{G_F^2\eta_{\rm EW}^2|V_{\rm CKM}|^2}{256\pi^3}
\frac{q^2}{M^2}
\left(1-\frac{m_\ell^2}{q^2}\right)^2,
\end{equation}
where $G_F$ is the Fermi constant, $V_{\rm CKM}$ is the
relevant CKM matrix element, and $\eta_{\rm EW}$ denotes
the electroweak correction factor~\cite{Sirlin1982,FNALMILC2015}.
The numerical inputs are specified in Sec.~\ref{sec:results}.
The helicity amplitudes are related to the form factors by
\begin{equation}
H_0(q^2)
=
\frac{2M|\bm p|}{\sqrt{q^2}}\,f_+(q^2),
\qquad
H_t(q^2)
=
\frac{M^2-M'^2}{\sqrt{q^2}}\,f_0(q^2).
\end{equation}

Substituting these amplitudes into Eq.~\eqref{eq:helicitywidth}
gives the expression used for the numerical integration,
\begin{align}
\frac{d\Gamma_\ell}{dq^2}
={}&
\frac{G_F^2\eta_{\rm EW}^2|V_{\rm CKM}|^2}{192\pi^3M^3}
\sqrt{\lambda(q^2)}
\left(1-\frac{m_\ell^2}{q^2}\right)^2
\nonumber\\
&\times
\left[
\lambda(q^2)
\left(1+\frac{m_\ell^2}{2q^2}\right)|f_+(q^2)|^2
+\frac{3m_\ell^2}{2q^2}
(M^2-M'^2)^2|f_0(q^2)|^2
\right],
\label{eq:physicalwidth}
\end{align}
where
$\lambda(q^2)=[(M+M')^2-q^2][(M-M')^2-q^2]$.
The physical decay interval is
$m_\ell^2\le q^2\le q^2_{\max}$, with
$q^2_{\max}=(M-M')^2$.
The total width and branching fraction are
\begin{equation}
\Gamma_\ell
=
\int_{m_\ell^2}^{q^2_{\max}}
\mathrm d q^2\,\frac{d\Gamma_\ell}{dq^2},
\qquad
\mathcal B_\ell=\frac{\tau_P}{\hbar}\,\Gamma_\ell,
\label{eq:totalwidth}
\end{equation}
where $\tau_P$ is the parent-meson lifetime.
The width vanishes when $m_\ell^2\ge q^2_{\max}$.
Physical meson and lepton masses are used throughout
the decay kinematics and phase-space integration.

The recoil variable is defined by
\begin{equation}
w=v\cdot v'
=\frac{M^2+M'^2-q^2}{2MM'},
\end{equation}
where $v=P/M$ and $v'=P'/M'$ are the initial and final
meson four-velocities.
Zero recoil corresponds to $w=1$ and $q^2=q^2_{\max}$.
For a given lepton mode, maximum recoil occurs at
$q^2=m_\ell^2$.
The differential width in terms of $w$ is
\begin{equation}
\frac{d\Gamma_\ell}{dw}
=
2MM'\frac{d\Gamma_\ell}{dq^2},
\qquad
1\le w\le\frac{M^2+M'^2-m_\ell^2}{2MM'}.
\label{eq:recoilwidth}
\end{equation}
The factor $2MM'$ is the absolute Jacobian
$|\mathrm d q^2/\mathrm dw|$.
For the form factors, the maximum-recoil point is
$q^2=0$, corresponding to
$w_{\max}=(M^2+M'^2)/(2MM')$.
This is also the maximum recoil of the decay in the
massless-lepton limit.

\section{Results and Discussion}\label{sec:results}
\subsection{Model inputs}
\label{subsec:model-inputs}
We consider the five ground-state transitions
$B\to D$, $B_s\to D_s$, $B_c\to\eta_c$,
$B_c\to B$, and $B_c\to B_s$ using the five
spectroscopy-based parameter sets in Table~\ref{input}.
In $B_c\to B$, the daughter is $B_d^0$. Charge-conjugate transitions
are understood when relating the quark-current convention of
Eq.~\eqref{eq:Jdef} to the charge channels listed below.
The $B^+\to\bar D^0$ rates are evaluated with the
charged-channel physical masses and lifetime, using
the same isospin-symmetric constituent wave functions
as for the neutral channel.
The physical masses below are in GeV~\cite{PDG2020}.
\begin{align}
&M_{B^0}=5.2797,\quad M_{B^+}=5.2793,\quad
 M_{D^-}=1.8697,\quad M_{\bar D^0}=1.8648,
\nonumber\\
&M_{B_s}=5.3669,\quad M_{D_s}=1.9684,\quad
 M_{B_c}=6.2749,\quad M_{\eta_c}=2.9839,
\nonumber\\
&m_e=0.000511,\qquad
 m_\mu=0.1057,\qquad m_\tau=1.7769.
\label{eq:externalinputs}
\end{align}
The parent-meson lifetimes are
$\tau_{B^0}=1.519~\mathrm{ps}$,
$\tau_{B^+}=1.638~\mathrm{ps}$, and
$\tau_{B_c}=0.510~\mathrm{ps}$~\cite{PDG2020}, and
$\tau_{B_s}=1.520~\mathrm{ps}$~\cite{HFLAVLifetimes2024}.
We use $G_F=1.166\times10^{-5}~\mathrm{GeV}^{-2}$
and
$\hbar\simeq6.582\times10^{-25}~\mathrm{GeV\,s}$~\cite{PDG2020}.
The CKM magnitudes are fixed to
$|V_{cb}|=0.0402$, $|V_{cs}|=0.975$, and $|V_{cd}|=0.221$.
The adopted $|V_{cb}|$ coincides with the central value
of the HPQCD determination in Ref.~\cite{HPQCD2015}, and
$|V_{cs}|$ and $|V_{cd}|$ are taken from Ref.~\cite{PDG2024}.

Short-distance electroweak effects are included through
the multiplicative amplitude factor $\eta_{\rm EW}=1.0066$,
corresponding to the Sirlin correction used in
semileptonic $b\to c$
analyses~\cite{Sirlin1982,FNALMILC2015}.
This factor enters the decay widths as $\eta_{\rm EW}^2$.
The same value is adopted for the charm-changing
$B_c\to B_{(s)}$ channels as a common normalization
choice in the present comparison.
Additional channel-dependent electromagnetic corrections
are not included.

\subsection{Form factors and decay rates}
\label{subsec:ff-decay-results}

For each transition, we compare the predictions from
the common model set
\begin{equation}
\mathcal E
=
\{\mbox{Log-Mix-P},\mbox{Lin-Mix-P},\mbox{Lin-Mix-NP},
  \mbox{HO-1S-P},\mbox{Lin-1S-P}\}.
\label{eq:ensemble}
\end{equation}
For an observable $O_a(z)$, where $a$ labels a model
and $z$ denotes $q^2$ or $w$, we define
\begin{equation}
\begin{aligned}
\overline O(z)
&=\frac{1}{5}\sum_{a\in\mathcal E}O_a(z),\\
O_{\min}(z)
&=\min_{a\in\mathcal E}O_a(z),
\qquad
O_{\max}(z)=\max_{a\in\mathcal E}O_a(z).
\end{aligned}
\label{eq:globalband}
\end{equation}
The central curves show the equal-weight means
$\overline O(z)$, and the shaded bands span
$[O_{\min}(z),O_{\max}(z)]$.
All five implementations contribute at each point
without fitting or reweighting to experimental or
lattice results.
The bands describe the spread among the selected models,
not statistical confidence intervals.

Differential widths and lepton-flavor ratios are
calculated separately in each model before their means
and ranges are formed.
Likewise, branching-fraction ranges are obtained from
the model-specific integrated rates, rather than by
integrating the pointwise upper and lower band edges.
Since covariance matrices for the adopted model parameters
are unavailable, no statistical parameter uncertainties
are assigned. The quoted values also do not include
uncertainties from the CKM elements, lifetimes, or
radiative corrections. 

Table~\ref{tab:FFendpoint} lists the direct-overlap form factors
at maximum recoil, $q^2=0$, and zero recoil,
$q^2=q^2_{\max}=(M-M')^2$, for all five implementations.
The scalar form factor is constructed from Eq.~\eqref{eq:f0def}.
It satisfies $f_0(0)=f_+(0)$ and
\begin{equation}
f_0(q^2_{\max})=f_+(q^2_{\max})+
\frac{M-M'}{M+M'}f_-(q^2_{\max}).
\label{eq:endpointidentity}
\end{equation}

\begin{table*}[t]
\centering\footnotesize
\caption{Direct-overlap endpoint form factors at $q^2=0$
and $q^2_{\max}=(M-M')^2$. The independently evaluated
transverse- and minus-current extractions yield the same
$f_-$ within numerical accuracy.}
\label{tab:FFendpoint}
\setlength{\tabcolsep}{6.5pt}
\renewcommand{\arraystretch}{1.5}
\begin{tabular}{llrrrrrrrrrr}
\hline\hline
Transition & FF & \multicolumn{2}{c}{\mbox{Log-Mix-P}} & \multicolumn{2}{c}{\mbox{Lin-Mix-P}} & \multicolumn{2}{c}{\mbox{Lin-Mix-NP}} & \multicolumn{2}{c}{\mbox{HO-1S-P}} & \multicolumn{2}{c}{\mbox{Lin-1S-P}}\\ & & $0$ & $q^2_{\max}$ & $0$ & $q^2_{\max}$ & $0$ & $q^2_{\max}$ & $0$ & $q^2_{\max}$ & $0$ & $q^2_{\max}$\\\hline
$B\to D$ & $f_+$ & 0.6881 & 1.1078 & 0.6938 & 1.1211 & 0.6381 & 1.1238 & 0.6969 & 1.1209 & 0.7157 & 1.1235\\
 & $f_-$ & -0.3106 & -0.5062 & -0.3254 & -0.5329 & -0.2919 & -0.5378 & -0.3190 & -0.5142 & -0.3298 & -0.5230\\
 & $f_0$ & 0.6881 & 0.8663 & 0.6938 & 0.8670 & 0.6381 & 0.8672 & 0.6969 & 0.8756 & 0.7157 & 0.8740\\
\hline
$B_s\to D_s$ & $f_+$ & 0.6815 & 1.1100 & 0.6725 & 1.1211 & 0.6259 & 1.1205 & 0.6593 & 1.1144 & 0.6832 & 1.1238\\
 & $f_-$ & -0.3177 & -0.5245 & -0.3241 & -0.5472 & -0.2899 & -0.5380 & -0.3041 & -0.5135 & -0.3235 & -0.5373\\
 & $f_0$ & 0.6815 & 0.8670 & 0.6725 & 0.8675 & 0.6259 & 0.8712 & 0.6593 & 0.8765 & 0.6832 & 0.8749\\
\hline
$B_c\to\eta_c$ & $f_+$ & 0.5800 & 1.0734 & 0.5281 & 1.0674 & 0.5212 & 1.0616 & 0.5458 & 1.0304 & 0.4817 & 1.0707\\
 & $f_-$ & -0.3136 & -0.6040 & -0.2858 & -0.5914 & -0.2689 & -0.5939 & -0.2452 & -0.4588 & -0.2665 & -0.6036\\
 & $f_0$ & 0.5800 & 0.8587 & 0.5281 & 0.8572 & 0.5212 & 0.8505 & 0.5458 & 0.8673 & 0.4817 & 0.8562\\
\hline
$B_c\to B$ & $f_+$ & 0.6340 & 0.8901 & 0.5572 & 0.8396 & 0.5082 & 0.7554 & 0.4282 & 0.6458 & 0.4642 & 0.7272\\
 & $f_-$ & -1.5468 & -2.5915 & -1.3975 & -2.4686 & -1.0420 & -1.9183 & -0.6755 & -1.1694 & -1.3069 & -2.3050\\
 & $f_0$ & 0.6340 & 0.6669 & 0.5572 & 0.6270 & 0.5082 & 0.5902 & 0.4282 & 0.5451 & 0.4642 & 0.5287\\
\hline
$B_c\to B_s$ & $f_+$ & 0.7227 & 0.9351 & 0.6617 & 0.9016 & 0.5691 & 0.7790 & 0.5736 & 0.7701 & 0.5695 & 0.8015\\
 & $f_-$ & -1.3702 & -2.0716 & -1.2461 & -1.9418 & -0.9100 & -1.5484 & -0.6103 & -0.9106 & -1.2369 & -1.9180\\
 & $f_0$ & 0.7227 & 0.7735 & 0.6617 & 0.7501 & 0.5691 & 0.6582 & 0.5736 & 0.6990 & 0.5695 & 0.6519\\
\hline\hline
\end{tabular}
\end{table*}

\begin{figure*}[t]
\centering
\includegraphics[width=0.4938\linewidth]{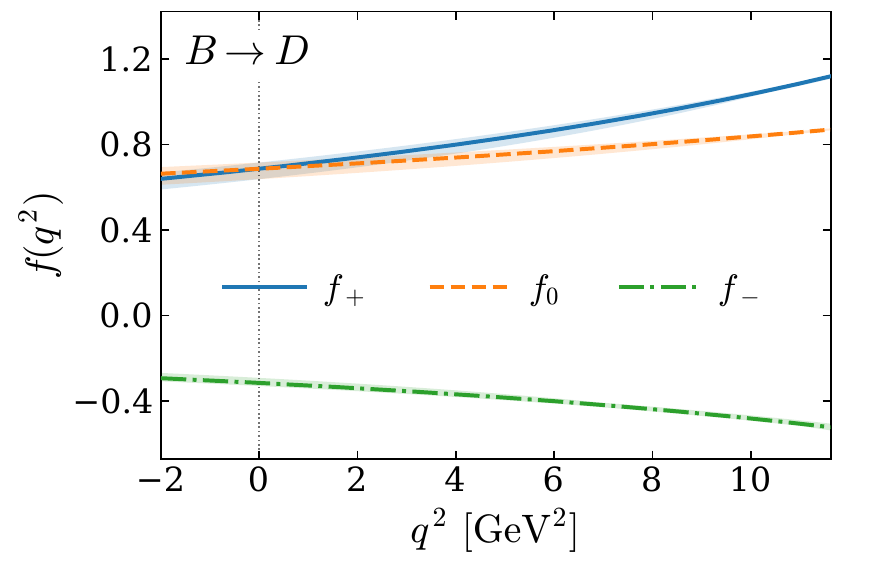}\\[0.5em]
\includegraphics[width=0.4938\linewidth]{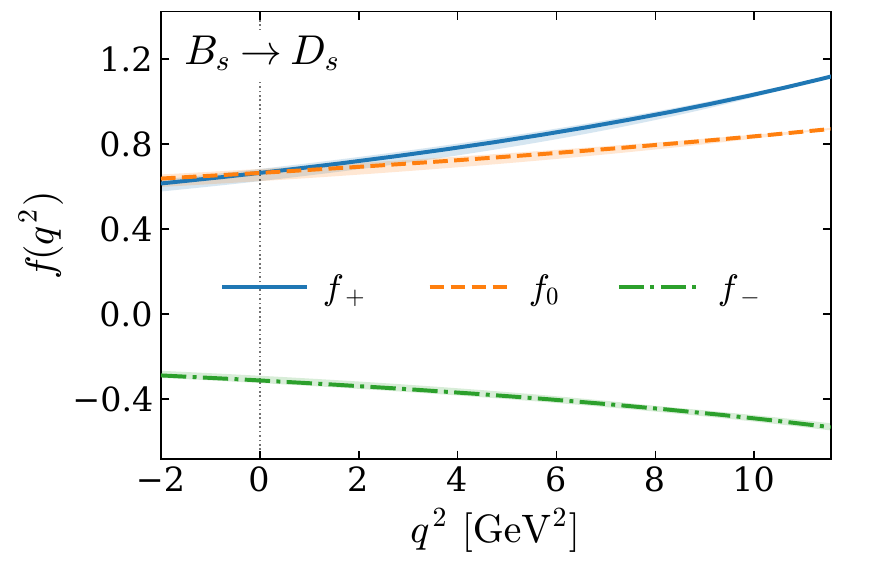}\hfill\includegraphics[width=0.4938\linewidth]{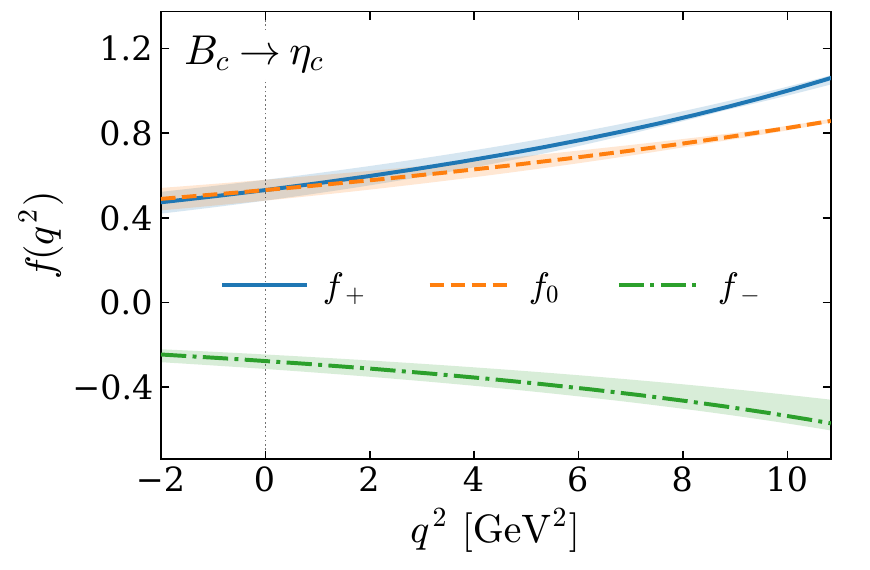}\\[0.5em]
\includegraphics[width=0.4938\linewidth]{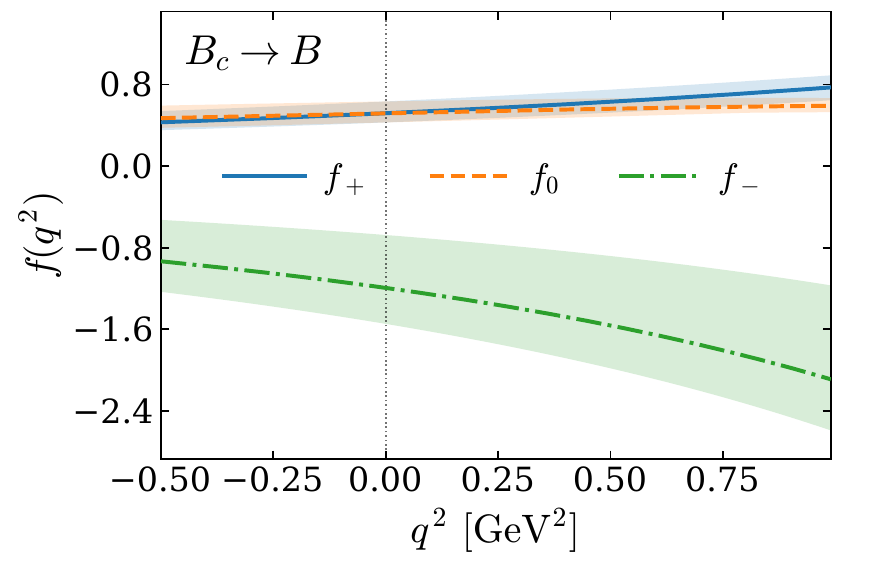}\hfill\includegraphics[width=0.4938\linewidth]{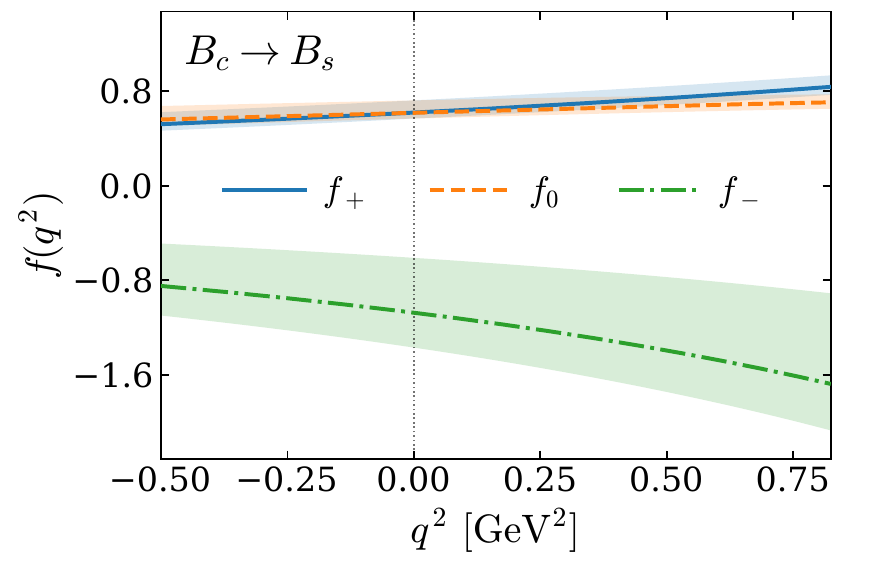}
\caption{Five-model means and minimum--maximum bands for
$f_+(q^2)$, $f_0(q^2)$, and $f_-(q^2)$, defined in
Eq.~\eqref{eq:globalband}. The solid, dashed, and dash-dotted curves
represent $f_+$, $f_0$, and $f_-$, respectively.
The two independently integrated current-component extractions of
$f_-$ agree within numerical accuracy and are shown only once.}
\label{formfactors1}
\end{figure*}
Figure~\ref{formfactors1} shows the corresponding $q^2$ dependence
in the spacelike and physical timelike regions, with the model means
and bands defined in Eq.~\eqref{eq:globalband}.
The transverse- and minus-current extractions of $f_-$ are evaluated
independently for every model and transition. They agree within
our numerical accuracy and are therefore displayed as a single
curve and band.

The $f_+$ and $f_0$ bands are particularly narrow for
$B\to D$ and $B_s\to D_s$ near zero recoil.
For $B_c\to\eta_c$, these form factors also become closely grouped
near $q^2_{\max}$, whereas $f_-$ retains a more pronounced model
spread. The largest variation in $f_-$ occurs in the charm-changing
$B_c\to B_{(s)}$ channels.
At zero recoil, $f_-$ ranges from $-2.5915$ to $-1.1694$ for
$B_c\to B$ and from $-2.0716$ to $-0.9106$ for $B_c\to B_s$.
The corresponding $f_+$ and $f_0$ values also show appreciable model
dependence, as listed in Table~\ref{tab:FFendpoint}, although their
absolute spreads are smaller than that of $f_-$.
This pattern is qualitatively consistent with the different
spin and momentum weights in the current projections.
The plus-current extraction of $f_+$ involves only the
helicity-nonflip kernel, whereas $f_-$ combines this overlap with a
transverse or minus current containing both nonflip and flip terms.
The additional constituent-mass and momentum dependence in
Eq.~\eqref{eq:projectedkernels} allows changes in the radial wave
functions and constituent parameters to affect $f_-$ differently
from $f_+$.
The larger spread in $f_-$ therefore probes model sensitivity within
the common current prescription, rather than a discrepancy between
current-component extractions.

For $B_c\to B_{(s)}$, the overlap connects a heavy--heavy initial
state to a heavy--light final state with the same bottom spectator.
The width ratios $\beta_{qb}/\beta_{cb}$ and
$\beta_{sb}/\beta_{cb}$ in Table~\ref{input} differ appreciably
among the implementations.
For \mbox{Log-Mix-P}, \mbox{Lin-Mix-P}, and \mbox{Lin-Mix-NP},
these ratios are $0.686$, $0.645$, and $0.636$ for $B_c\to B$,
and $0.738$, $0.701$, and $0.664$ for $B_c\to B_s$.
The values of $f_+(0)$ decrease in the same order in both channels.
The \mbox{HO-1S-P} implementation has the largest initial width,
$\beta_{cb}=1.035~\mathrm{GeV}$, and the smallest
$\beta_{qb}/\beta_{cb}=0.479$.
It gives the smallest $f_+(0)$ for $B_c\to B$ and the smallest
$|f_-|$ at both endpoints in the two $B_c\to B_{(s)}$ channels.
These correlations suggest sensitivity to the initial--final radial
overlap, but do not isolate its contribution from changes in the
constituent masses and basis mixing. For example, \mbox{Lin-1S-P}
does not follow the ordering of the mixed-basis implementations.
The bands consequently reflect differences among the complete
spectroscopic implementations, not the confinement form alone.

A complementary kinematic interpretation follows from
Eq.~\eqref{eq:f0def}. In the physical form-factor interval,
the coefficient multiplying $f_-$ obeys
\begin{equation*}
0\le\frac{q^2}{M^2-M'^2}
\le\frac{M-M'}{M+M'}.
\end{equation*}
The upper bound is approximately $0.0861$ for $B_c\to B$ and
$0.0780$ for $B_c\to B_s$.
Thus, a variation of $f_-$ contributes to $f_0$ with a small
kinematic weight, and the accompanying changes in $f_+$ and $f_-$
can partially compensate in their scalar combination.
A broad $f_-$ band is therefore compatible with narrower
$f_+$ and $f_0$ bands.
At $q^2=0$, the coefficient of $f_-$ vanishes, so that
$f_0(0)=f_+(0)$ by construction. The information carried by $f_-(0)$
then appears only in the slope difference,
$f_-(0)=(M^2-M'^2)[f_0'(0)-f_+'(0)]$, obtained by differentiating
Eq.~\eqref{eq:f0def}, with primes denoting derivatives with respect
to $q^2$.
For applications that require the form factors at arbitrary $q^2$,
Appendix~\ref{app:auxiliary} provides a compact auxiliary rational
parametrization of the directly calculated results. It is not used
to obtain the results presented in this section.

\begin{figure*}[t]
\centering
\includegraphics[width=0.48\linewidth]{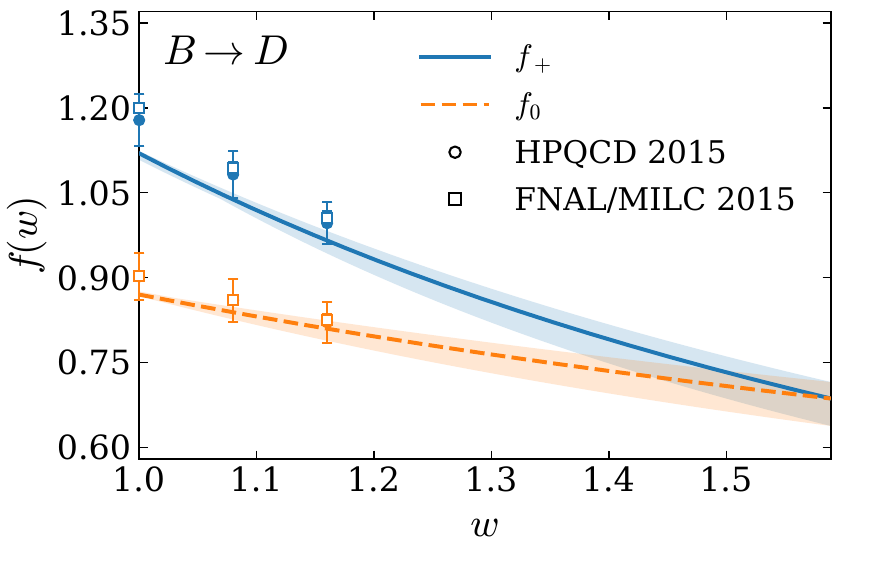}\\[0.5em]
\includegraphics[width=0.48\linewidth]{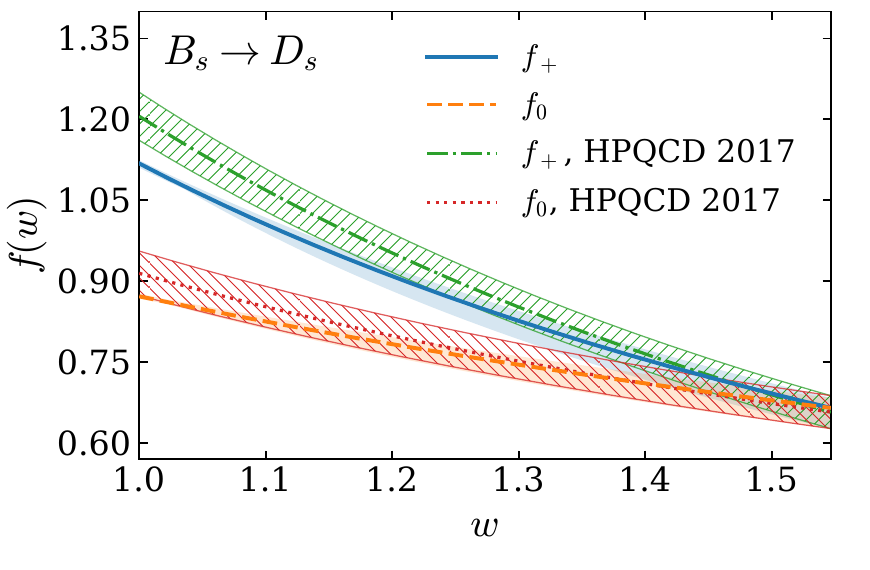}\hfill\includegraphics[width=0.485\linewidth]{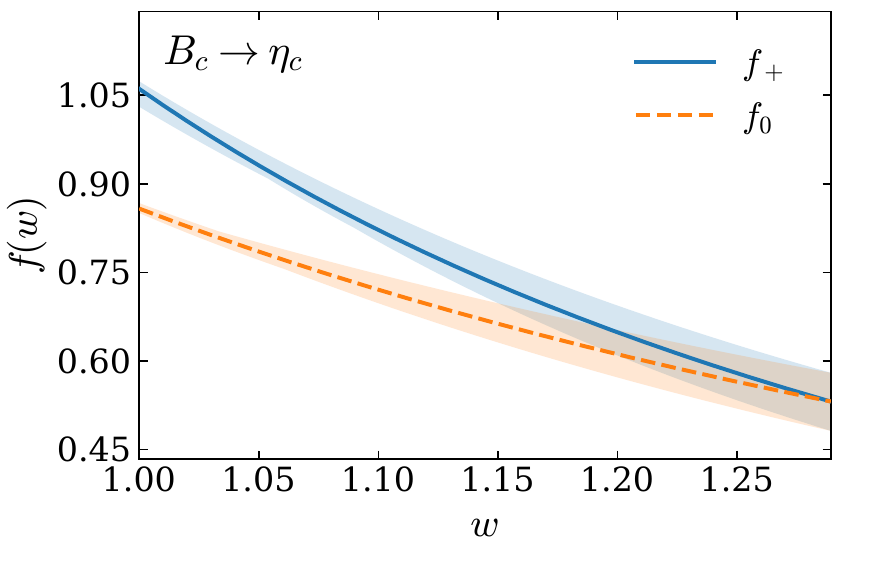}\\[0.5em]
\includegraphics[width=0.48\linewidth]{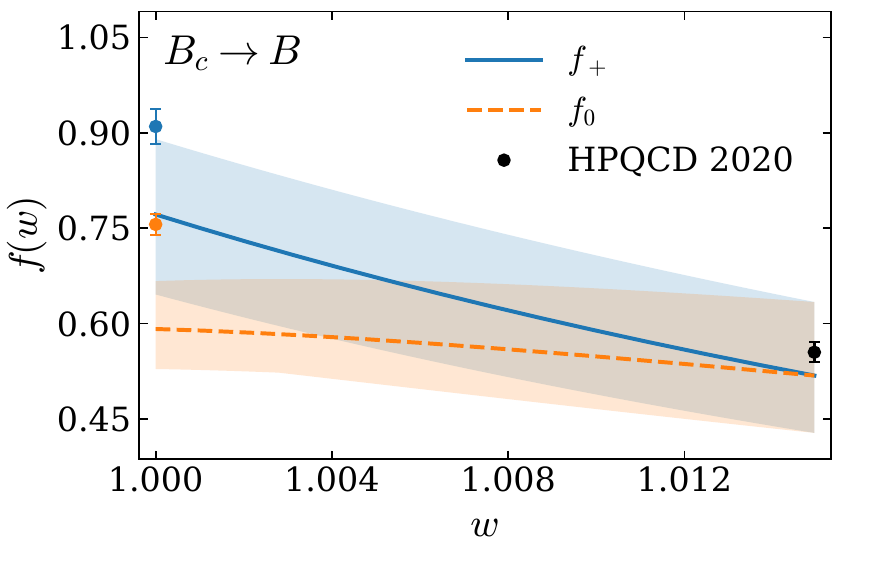}\hfill\includegraphics[width=0.485\linewidth]{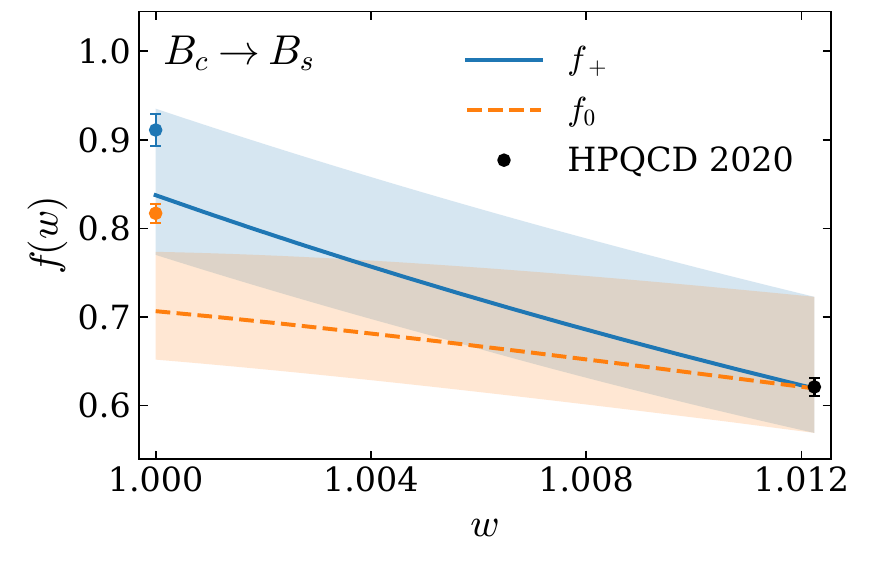}
\caption{Five-model means and ranges for $f_+(w)$ and $f_0(w)$.
Lattice results are shown for $B\to D$, $B_s\to D_s$, and
$B_c\to B_{d,s}$~\cite{HPQCD2015,FNALMILC2015,HPQCD2017,Cooper2020}.
The $B_c\to\eta_c$ panel shows the model predictions alone.
In the $B_c\to B_{(s)}$ panels, the colored points at $w=1$ give
$f_+(q^2_{\max})$ and $f_0(q^2_{\max})$, and the black point at maximum
recoil gives the common value $f_+(0)=f_0(0)$.}
\label{fofprecoil}
\end{figure*}

Figure~\ref{fofprecoil} presents the $f_+$ and $f_0$ results
of Fig.~\ref{formfactors1} as functions of the recoil
variable $w$, together with available lattice results.
In the $B\to D$ panel, circles and squares denote the HPQCD and
Fermilab/MILC values~\cite{HPQCD2015,FNALMILC2015} at $w=1$, $1.08$,
and $1.16$, as tabulated in Ref.~\cite{Belle2015}.
The overall recoil dependence is qualitatively similar,
but our zero-recoil $f_+$ values, approximately
$1.108$--$1.124$, lie below the Fermilab/MILC
value $1.1994(95)$ and about $1.2$--$1.5$ standard deviations below
the HPQCD value $1.178(46)$.

For $B_s\to D_s$, the green and red hatched bands show
the HPQCD results for $f_+$ and $f_0$, respectively,
with pointwise one-standard-deviation
uncertainties~\cite{HPQCD2017}.
Our model bands overlap the corresponding HPQCD bands
near maximum recoil.
At zero recoil, our $f_+$ band lies below the HPQCD
band, whereas the $f_0$ bands partially overlap.

For $B_c\to B$ and $B_c\to B_s$, the points show the
HPQCD 2020 endpoint form factors, as corrected in its
erratum~\cite{Cooper2020}.
The black points at maximum recoil denote the common
value $f_+(0)=f_0(0)$, while the blue and orange points
at $w=1$ give $f_+(q^2_{\max})$ and $f_0(q^2_{\max})$,
respectively.
The maximum-recoil lattice values, $0.555(16)$ for
$B_c\to B$ and $0.621(10)$ for $B_c\to B_s$,
lie within the corresponding five-model bands.

At zero recoil, the vector and scalar comparisons differ.
For $B_c\to B_s$, the lattice value
$f_+(q^2_{\max})=0.911(18)$ lies within our model band.
For $B_c\to B$, the lattice central value of $0.910$
is above all five model predictions, but its quoted
one-standard-deviation interval, $0.910\pm0.028$,
overlaps the upper part of the model band.
In contrast, all five predictions for $f_0(q^2_{\max})$
lie below the lattice one-standard-deviation intervals
in both channels, with HPQCD values of $0.756(16)$
for $B_c\to B$ and $0.817(11)$ for
$B_c\to B_s$~\cite{Cooper2020}.

Table~4 of Ref.~\cite{ThejusMary2025} reports type-II covariant LFQM
results for $B_c\to B$ and $B_c\to B_s$ using a three-parameter
rational form (T2A) and a $z$ expansion (T2B).
At maximum recoil, the T2A and T2B central values for $f_+(0)=f_0(0)$
are $0.60$ and $0.62$ for $B_c\to B$, and $0.71$ and $0.72$ for
$B_c\to B_s$. All four lie within the corresponding five-model ranges.
At zero recoil, the corresponding central values of $f_0(q^2_{\max})$
are $0.79$ and $0.86$ for $B_c\to B$, and $0.90$ and
$0.95$ for $B_c\to B_s$. These values exceed the corresponding HPQCD
central values, whereas our largest predictions, $0.6669$ and $0.7735$,
respectively, lie below them.
The zero-recoil values in that study are extrapolated from fits to
spacelike form factors. Here we integrate the analytically continued
overlaps directly. The two calculations also use different constituent
masses and radial wave functions.

For $B\to D$ and $B_s\to D_s$, we use the conventional
finite-mass form factor~\cite{HPQCD2015,HPQCD2017}
\begin{equation}
\mathcal G_{P\to P'}(w)
=
\frac{2\sqrt{MM'}}{M+M'}\,f_+(w),
\label{eq:IWeffective}
\end{equation}
where $M$ and $M'$ are the physical meson masses.
This combination is related to the Isgur--Wise function
in the heavy-quark limit~\cite{Neubert1993}, but its
normalization is not constrained to unity at the physical
masses.

The zero-recoil value $\mathcal G(1)$ specifies the
normalization, while the normalized slope $\rho^2(1)$
characterizes the relative recoil dependence near $w=1$.
They are given by
\begin{align}
\mathcal G(1)
&=
\frac{2\sqrt{MM'}}{M+M'}\,f_+(q^2_{\max}),
\label{eq:Gzero}\\
\rho^2(1)
&=
-\left.
\frac{1}{\mathcal G(w)}
\frac{d\mathcal G(w)}{dw}
\right|_{w=1}
\nonumber\\
&=
\frac{2MM'}{f_+(q^2_{\max})}
\left.
\frac{df_+(q^2)}{dq^2}
\right|_{q^2=q^2_{\max}},
\label{eq:rhozero}
\end{align}
where $q^2_{\max}=(M-M')^2$ and $dq^2/dw=-2MM'$.
With this convention,
$\mathcal G(w)/\mathcal G(1)
=1-\rho^2(1)(w-1)+\mathcal O((w-1)^2)$.
We evaluate $f_+(q^2_{\max})$ directly from the continued
overlap. For the endpoint derivative, we differentiate the
full integrand with respect to $q^2$ at fixed $x$ and
$\bm\ell_\perp$ before integration.

\begin{table*}[t]
\centering\small
\caption{LFQM predictions for the zero-recoil normalization
$\mathcal G(1)$ and normalized slope $\rho^2(1)$
for $B\to D$ and $B_s\to D_s$.
Our results are compared with the available experimental
results~\cite{BaBar2024,BelleII2025Vcb} and lattice QCD
determinations~\cite{HPQCD2015,HPQCD2017}.
The BaBar values are from its CLN fit, which includes
Fermilab/MILC lattice and Belle constraints.}
\label{tab:zeroRecoilComparison}
\setlength{\tabcolsep}{16pt}
\renewcommand{\arraystretch}{1.3}
\begin{tabular}{lcccc}
\hline\hline
 & \multicolumn{2}{c}{$B\to D$}
 & \multicolumn{2}{c}{$B_s\to D_s$}\\
Source & $\mathcal G(1)$ & $\rho^2(1)$
       & $\mathcal G(1)$ & $\rho^2(1)$\\
\hline
\mbox{Log-Mix-P}
 & 0.974 & 0.946 & 0.984 & 1.031\\
\mbox{Lin-Mix-P}
 & 0.985 & 0.944 & 0.994 & 1.064\\
\mbox{Lin-Mix-NP}
 & 0.988 & 1.164 & 0.993 & 1.301\\
\mbox{HO-1S-P}
 & 0.985 & 0.918 & 0.988 & 1.070\\
\mbox{Lin-1S-P}
 & 0.987 & 0.872 & 0.996 & 1.016\\
\hline
\rule{0pt}{3.0ex}Global mean and spread
 & $0.984^{+0.004}_{-0.010}$
 & $0.969^{+0.195}_{-0.097}$
 & $0.991^{+0.005}_{-0.007}$
 & $1.096^{+0.204}_{-0.080}$\\
\hline
HPQCD 2015~\cite{HPQCD2015}
 & $1.035(40)$ & $1.119(71)$ & -- & --\\
BaBar 2024 (CLN)~\cite{BaBar2024}
 & $1.056(8)$ & $1.155(23)$ & -- & --\\
Belle II 2025 (CLN)~\cite{BelleII2025Vcb}
 & -- & $1.09(6)$ & -- & --\\
HPQCD 2017~\cite{HPQCD2017}
 & -- & -- & $1.068(40)$ & $1.244(76)$\\
\hline\hline
\end{tabular}
\end{table*}
Table~\ref{tab:zeroRecoilComparison} compares our predictions
for $\mathcal G(1)$ and $\rho^2(1)$ with the HPQCD lattice
results~\cite{HPQCD2015,HPQCD2017}.
For $B\to D$, the table also lists the normalization and slope
from the BaBar Caprini--Lellouch--Neubert (CLN)
fit~\cite{BaBar2024} and the slope from the Belle~II CLN
fit~\cite{BelleII2025Vcb}.
Our five-model means for $\mathcal G(1)$, $0.984$ for
$B\to D$ and $0.991$ for $B_s\to D_s$, are approximately
$4.9\%$ and $7.2\%$ below the respective HPQCD central values.
In each channel, even the largest model prediction for
$\mathcal G(1)$ lies below the lower edge of the quoted
HPQCD one-standard-deviation interval.

In contrast, the HPQCD central values of $\rho^2(1)$ lie
within the five-model ranges of $0.872$--$1.164$ for
$B\to D$ and $1.016$--$1.301$ for $B_s\to D_s$.
The \mbox{Lin-Mix-NP} implementation gives the largest
slope in both channels.
For $B\to D$, the central slopes from the BaBar and
Belle~II CLN fits, $1.155$~\cite{BaBar2024} and
$1.09$~\cite{BelleII2025Vcb}, respectively, also lie
within the model range and above our five-model mean
of $0.969$.

Using the Godfrey--Isgur spectroscopic model~\cite{Godfrey1985}
in the BT framework, Mor\'enas \textit{et al.}~\cite{Morenas1997}
obtained heavy-quark-limit Isgur--Wise slopes of $1.026$
and $1.15$ for nonstrange and strange spectators, respectively.
All five finite-mass implementations show the same ordering,
with a larger zero-recoil slope for $B_s\to D_s$ than
for $B\to D$.

\begin{figure*}[t]
\centering
\includegraphics[width=0.485\linewidth]{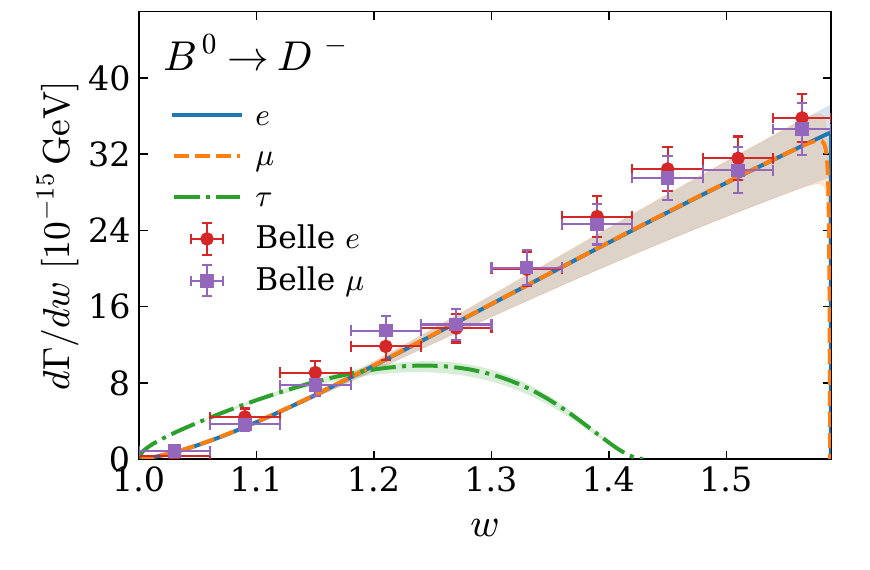}\hfill
\includegraphics[width=0.485\linewidth]{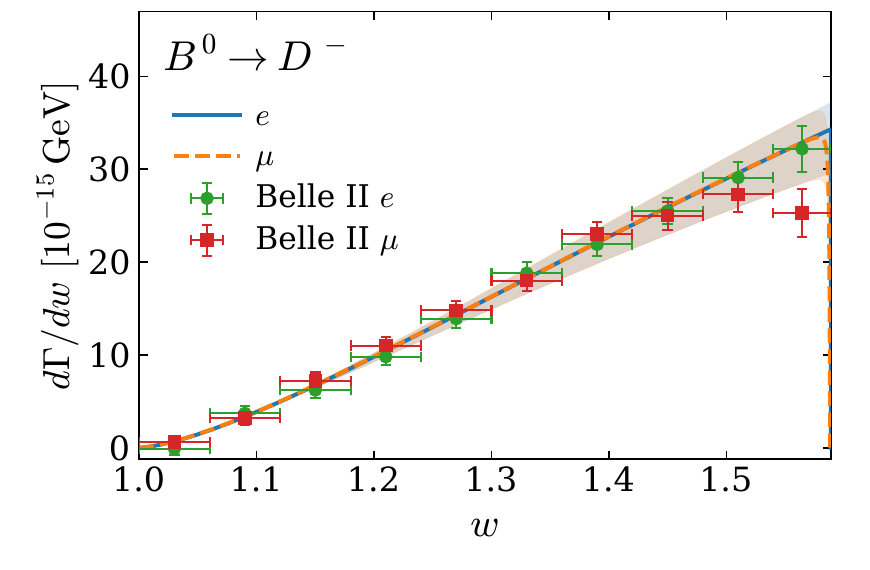}\\[0.5em]
\includegraphics[width=0.485\linewidth]{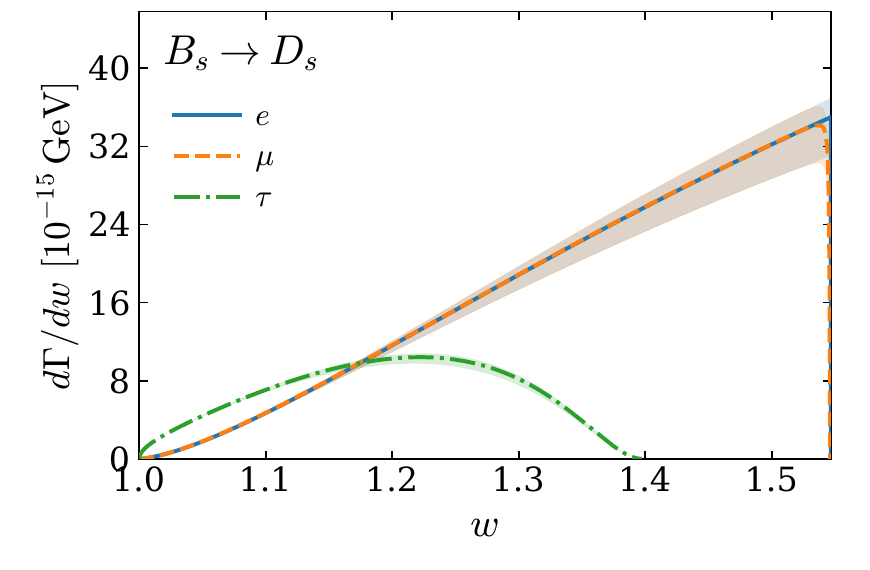}\hfill\includegraphics[width=0.485\linewidth]{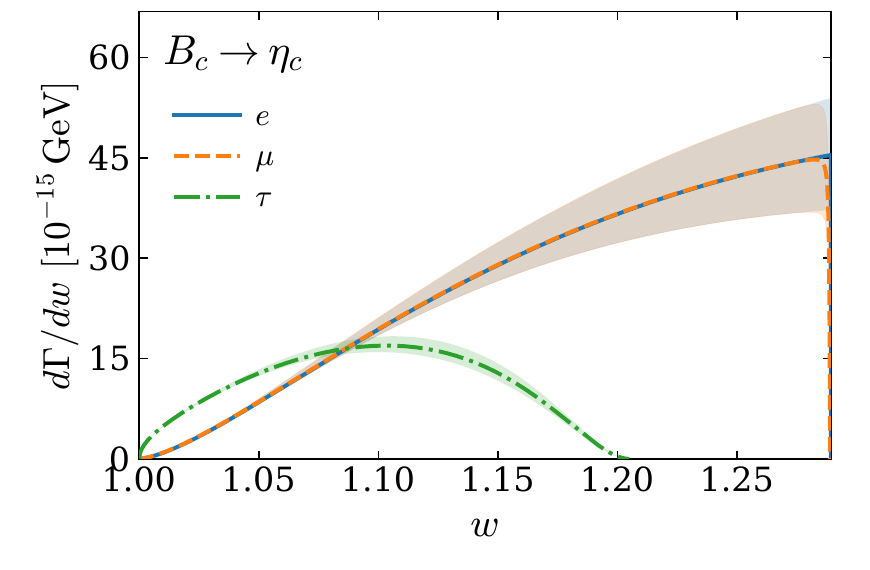}\\[0.5em]
\includegraphics[width=0.485\linewidth]{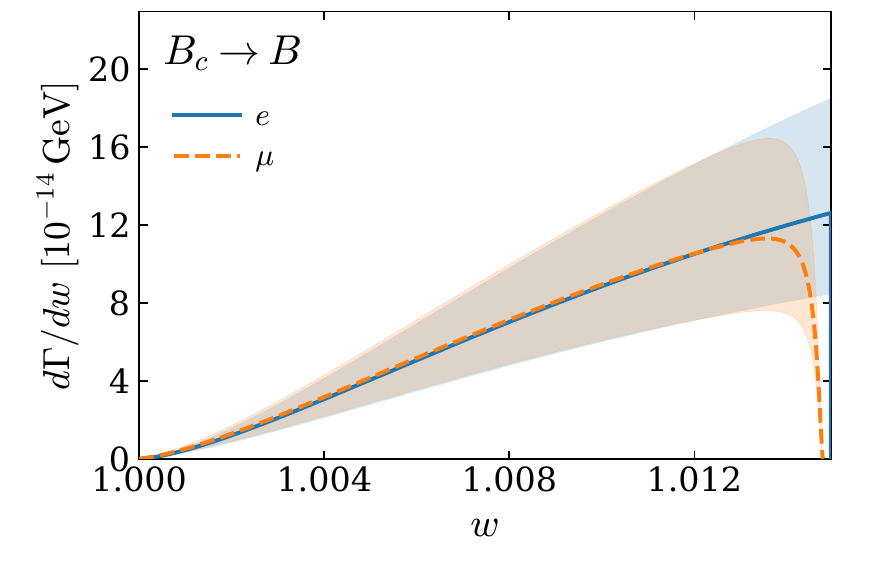}\hfill\includegraphics[width=0.485\linewidth]{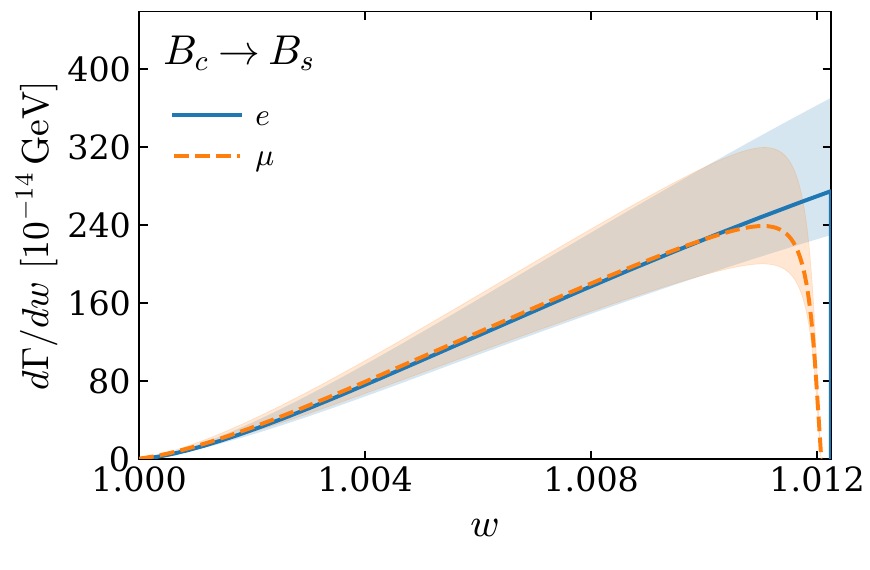}
\caption{Five-model means and minimum--maximum bands for
$d\Gamma/dw$. The first row compares the neutral-channel $B\to D$
predictions with the Belle~\cite{Belle2015} (left)
and Belle II~\cite{BelleII2025Vcb} (right) measurements.
The Belle II points are the Coulomb-corrected neutral-channel electron
and muon subsamples in Table~III of Ref.~\cite{BelleII2025Vcb}.}
\label{diffdecay}
\end{figure*}
\begin{figure*}[t]
\centering
\includegraphics[width=0.485\linewidth]{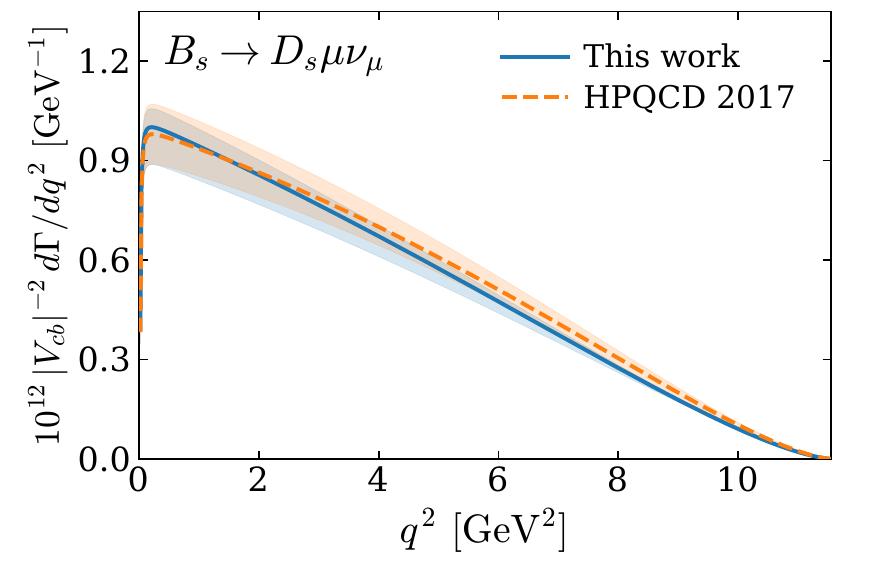}\hfill
\includegraphics[width=0.485\linewidth]{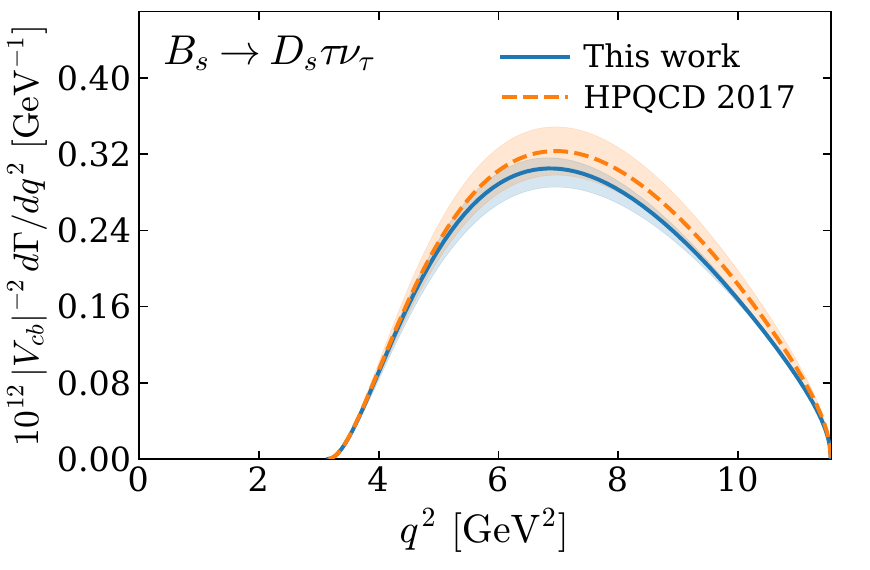}
\caption{Comparison of our LFQM predictions for the
$B_s\to D_s\mu\nu$ and $B_s\to D_s\tau\nu$
differential decay widths with the HPQCD lattice QCD
results~\cite{HPQCD2017}.}
\label{fig:BsDsDiffComp}
\end{figure*}
Figure~\ref{diffdecay} shows $d\Gamma/dw$ for the electron,
muon, and kinematically allowed $\tau$ modes. For each mode,
the curve gives the five-model mean and the shaded band
spans the range of model predictions.
The electron and muon spectra nearly coincide over most of
the recoil range in $B_{(s)}\to D_{(s)}$ and $B_c\to\eta_c$.
In $B_c\to B_{(s)}$, the smaller energy release makes the
muon-mass suppression more pronounced near maximum recoil,
and the $\tau$ modes are kinematically forbidden.

The upper panels compare the same $B^0\to D^-\ell^+\nu_\ell$
predictions for $\ell=e,\mu$ with the Belle~\cite{Belle2015}
(left) and Belle~II~\cite{BelleII2025Vcb} (right) measurements.
The experimental points give the bin-averaged differential widths
$\Delta\Gamma_i/\Delta w_i$, while the curves show $d\Gamma/dw$
evaluated pointwise in $w$.

Figure~\ref{fig:BsDsDiffComp} compares the $B_s\to D_s$ muon
and $\tau$ spectra obtained from our form factors with those
obtained from the HPQCD lattice form factors~\cite{HPQCD2017}.
We plot $10^{12}|V_{cb}|^{-2}\,d\Gamma/dq^2$, using the same
$G_F$ and $\eta_{\rm EW}$ for both sets of spectra.
The muon spectra have similar shapes, with our five-model mean
slightly below the HPQCD central curve at intermediate and
large $q^2$. For the $\tau$ mode, the HPQCD central curve
also lies above our mean near the peak at
$q^2\simeq 7~\mathrm{GeV}^2$.

Table~\ref{t41} lists the branching fractions for the
semileptonic $B_{(s)}\to D_{(s)}$ and $B_c\to(\eta_c,B,B_s)$
decays. For $B^0\to D^-\ell^+\nu_\ell$ and
$B^+\to\bar D^0\ell^+\nu_\ell$, we average the electron
and muon branching fractions within each model.
The resulting five-model means and spreads are
$(2.186^{+0.144}_{-0.226})\%$ and
$(2.369^{+0.156}_{-0.244})\%$, respectively.
Belle~II reports $(2.06\pm0.05\pm0.10)\%$ for the neutral
channel and $(2.31\pm0.04\pm0.09)\%$ for the charged
channel~\cite{BelleII2025Vcb}, where the first uncertainty
is statistical and the second systematic.
Both measured central values lie within the corresponding
five-model ranges.

For the corresponding $\tau$ modes, the five-model results
are $(0.629^{+0.027}_{-0.040})\%$ for $B^0\to D^-\tau^+\nu_\tau$
and $(0.685^{+0.029}_{-0.043})\%$ for
$B^+\to\bar D^0\tau^+\nu_\tau$.
The 2024 PDG quotes $(0.99\pm0.21)\%$ and $(0.77\pm0.25)\%$
for the neutral and charged channels, respectively~\cite{PDG2024}.
All five charged-channel predictions lie within the quoted
one-standard-deviation interval. For the neutral channel,
even the largest model prediction, $0.656\%$, lies below
the lower edge of the corresponding interval, $0.78\%$.

The five-model predictions for
$\mathcal B(B_s^0\to D_s^-\mu^+\nu_\mu)$ range from
$1.934\%$ to $2.226\%$, with an equal-weight mean of
$2.125\%$. This mean lies within the quoted one-standard-deviation
interval of the 2024 PDG value, $(2.31\pm0.21)\%$~\cite{PDG2024}.
The PDG obtains this value by multiplying the LHCb
branching-fraction ratio by its reference value for
$\mathcal B(B^0\to D^-\ell^+\nu_\ell)$.
We calculate the ratio of the two muon-mode branching
fractions separately for each implementation and obtain
the five-model mean and spread
\begin{equation}
\frac{\mathcal B(B_s^0\to D_s^-\mu^+\nu_\mu)}
     {\mathcal B(B^0\to D^-\mu^+\nu_\mu)}
=0.975^{+0.033}_{-0.032}.
\label{eq34}
\end{equation}
The common factors $|V_{cb}|^2$ and $\eta_{\rm EW}^2$
cancel in this ratio, which retains the lifetime ratio
$\tau_{B_s^0}/\tau_{B^0}$, the phase-space differences, and
the spectator-flavor dependence of the form factors.
LHCb measures $1.09\pm0.05_{\rm stat}\pm0.06_{\rm syst}
\pm0.05_{\rm ext}$~\cite{LHCb2020}. Combining these
uncertainties in quadrature gives a lower one-standard-deviation
bound of $0.997$, just below the upper end of our model
range, $1.008$.

For $B_c\to\eta_c$, the electron and muon branching fractions
together span $0.511\%$ to $0.655\%$ across the five models.
The $\tau$ branching fractions range from $0.168\%$ to
$0.191\%$. The light-lepton branching fractions for
$B_c\to B_s$ and $B_c\to B$ show a larger relative spread
than those for $B_c\to\eta_c$, consistent with the stronger
model dependence of their form factors.

\begin{table*}[t]
\centering\footnotesize
\caption{Branching fractions (in percent) for the semileptonic
$B_{(s)}\to D_{(s)}$ and $B_c\to(\eta_c,B,B_s)$ decays.
For $B^0\to D^-$ and $B^+\to\bar D^0$, the entries
labeled $\ell$ denote the averages of the electron- and
muon-mode branching fractions, $\mathcal B_\ell=(\mathcal B_e+\mathcal B_\mu)/2$.}
\label{t41}
\setlength{\tabcolsep}{8pt}
\renewcommand{\arraystretch}{1.5}
\begin{tabular}{lrrrrrcc}
\hline\hline
Decay
& \mbox{Log-Mix-P}
& \mbox{Lin-Mix-P}
& \mbox{Lin-Mix-NP}
& \mbox{HO-1S-P}
& \mbox{Lin-1S-P}
& \shortstack{Global mean\\and spread}
& Experiment~\cite{PDG2024}\\
\hline
$B^0\to D^-\,\ell^+\nu_{\ell}$
 & 2.177 & 2.222 & 1.961 & 2.241 & 2.330
 & $2.186^{+0.144}_{-0.226}$ & $2.12\pm0.06$\\
$B^0\to D^-\,\tau^+\nu_{\tau}$
 & 0.625 & 0.634 & 0.590 & 0.642 & 0.656
 & $0.629^{+0.027}_{-0.040}$ & $0.99\pm0.21$\\
\hline
$B^+\to\bar D^0\,\ell^+\nu_{\ell}$
 & 2.359 & 2.407 & 2.125 & 2.429 & 2.525
 & $2.369^{+0.156}_{-0.244}$ & $2.21\pm0.06$\\
$B^+\to\bar D^0\,\tau^+\nu_{\tau}$
 & 0.680 & 0.690 & 0.642 & 0.699 & 0.714
 & $0.685^{+0.029}_{-0.043}$ & $0.77\pm0.25$\\
\hline
$B_s^0\to D_s^-\,e^+\nu_{e}$
 & 2.199 & 2.176 & 1.941 & 2.116 & 2.235
 & $2.133^{+0.101}_{-0.192}$ & $\cdots$\\
$B_s^0\to D_s^-\,\mu^+\nu_{\mu}$
 & 2.190 & 2.167 & 1.934 & 2.108 & 2.226
 & $2.125^{+0.101}_{-0.191}$ & $2.31\pm0.21$\\
$B_s^0\to D_s^-\,\tau^+\nu_{\tau}$
 & 0.629 & 0.628 & 0.586 & 0.623 & 0.643
 & $0.622^{+0.021}_{-0.035}$ & $\cdots$\\
\hline
$B_c^+\to\eta_c\,e^+\nu_{e}$
 & 0.655 & 0.577 & 0.550 & 0.589 & 0.512
 & $0.577^{+0.078}_{-0.065}$ & $\cdots$\\
$B_c^+\to\eta_c\,\mu^+\nu_{\mu}$
 & 0.653 & 0.575 & 0.549 & 0.587 & 0.511
 & $0.575^{+0.078}_{-0.064}$ & $\cdots$\\
$B_c^+\to\eta_c\,\tau^+\nu_{\tau}$
 & 0.191 & 0.178 & 0.170 & 0.181 & 0.168
 & $0.178^{+0.014}_{-0.010}$ & $\cdots$\\
\hline
$B_c^+\to B_s^0\,e^+\nu_{e}$
 & 1.627 & 1.408 & 1.041 & 1.048 & 1.066
 & $1.238^{+0.389}_{-0.197}$ & $\cdots$\\
$B_c^+\to B_s^0\,\mu^+\nu_{\mu}$
 & 1.530 & 1.328 & 0.982 & 0.992 & 1.005
 & $1.167^{+0.362}_{-0.185}$ & $\cdots$\\
\hline
$B_c^+\to B^0\,e^+\nu_{e}$
 & 0.105 & 0.084 & 0.069 & 0.050 & 0.060
 & $0.074^{+0.031}_{-0.024}$ & $\cdots$\\
$B_c^+\to B^0\,\mu^+\nu_{\mu}$
 & 0.099 & 0.080 & 0.066 & 0.048 & 0.057
 & $0.070^{+0.029}_{-0.022}$ & $\cdots$\\
\hline\hline
\end{tabular}
\end{table*}

For these charm-changing $B_c$ transitions, the electron-mode
branching fractions in Table~\ref{t41} correspond to decay rates
of approximately $(20.4\text{--}31.9)\times10^9~\mathrm{s}^{-1}$
for $B_c\to B_s$ and
$(0.98\text{--}2.05)\times10^9~\mathrm{s}^{-1}$
for $B_c\to B$.
The HPQCD results in Ref.~\cite{Cooper2020} are $26.25(90)(83)\times10^9~\mathrm{s}^{-1}$
and $1.650(61)(84)\times10^9~\mathrm{s}^{-1}$, respectively.
The first uncertainty arises from the CKM inputs and the
second from the lattice calculation.
These rates use $|V_{cs}|=0.997(17)$ and $|V_{cd}|=0.218(4)$,
whereas our fixed values are $0.975$ and $0.221$, respectively.
Their decay-rate formula contains no explicit
$\eta_{\rm EW}^2$ factor.

To remove these normalization factors, we define the reduced widths
\begin{equation}
\widehat\Gamma_q\equiv
\frac{\Gamma(B_c\to B_q e\nu_e)}
{|V_{cq}|^2\eta_{\rm EW}^2},\qquad q=s,d,
\label{eq:reducedBcwidth}
\end{equation}
where $B_d\equiv B$.
For the quoted lattice rates, $\eta_{\rm EW}=1$ in this
definition.
The lattice rates are evaluated in the massless-lepton
limit, so we compare them with our electron-mode results.
These reduced widths probe phase-space-weighted
form-factor integrals dominated by $|f_+(q^2)|^2$.

Our five-model means and spreads are
\begin{align*}
\widehat\Gamma_s
&=(2.520^{+0.792}_{-0.402})\times10^{10}~
\mathrm{s}^{-1},\\
\widehat\Gamma_d
&=(2.917^{+1.223}_{-0.943})\times10^{10}~
\mathrm{s}^{-1}.
\end{align*}
The corresponding HPQCD results are
$2.640(84)\times10^{10}~\mathrm{s}^{-1}$ and
$3.479(182)\times10^{10}~\mathrm{s}^{-1}$,
respectively~\cite{Cooper2020}.
The HPQCD central values exceed our means but lie within
the corresponding five-model ranges.

\begin{table*}[t]
\centering\small
\caption{Lepton-flavor ratios for the semileptonic
$B_{(s)}\to D_{(s)}$ and $B_c\to(\eta_c,B,B_s)$ decays.
We define $R_{\tau/e}=\Gamma_\tau/\Gamma_e$,
$R_{\tau/\mu}=\Gamma_\tau/\Gamma_\mu$,
$R=2\Gamma_\tau/(\Gamma_e+\Gamma_\mu)$, and
$R_{\mu/e}=\Gamma_\mu/\Gamma_e$.
A dash indicates a kinematically closed $\tau$ mode.}
\label{tab:ratios}
\setlength{\tabcolsep}{15pt}
\renewcommand{\arraystretch}{1.2}
\begin{tabular}{llrrrr}
\hline\hline
Transition & Model & $R_{\tau/e}$ & $R_{\tau/\mu}$
 & $R$ & $R_{\mu/e}$\\
\hline
$B\to D$ & \mbox{Log-Mix-P}
 & 0.286 & 0.288 & 0.287 & 0.996\\
 & \mbox{Lin-Mix-P}
 & 0.285 & 0.286 & 0.285 & 0.996\\
 & \mbox{Lin-Mix-NP}
 & 0.300 & 0.301 & 0.301 & 0.996\\
 & \mbox{HO-1S-P}
 & 0.286 & 0.287 & 0.287 & 0.996\\
 & \mbox{Lin-1S-P}
 & 0.281 & 0.282 & 0.282 & 0.996\\
\cline{2-6}
 & \rule{0pt}{3.0ex}Global mean and spread
 & $0.288^{+0.012}_{-0.007}$
 & $0.289^{+0.012}_{-0.007}$
 & $0.288^{+0.012}_{-0.007}$
 & $0.9959^{+0.0003}_{-0.0001}$\\
\hline
$B_s\to D_s$ & \mbox{Log-Mix-P}
 & 0.286 & 0.287 & 0.287 & 0.996\\
 & \mbox{Lin-Mix-P}
 & 0.289 & 0.290 & 0.289 & 0.996\\
 & \mbox{Lin-Mix-NP}
 & 0.302 & 0.303 & 0.303 & 0.996\\
 & \mbox{HO-1S-P}
 & 0.295 & 0.296 & 0.295 & 0.996\\
 & \mbox{Lin-1S-P}
 & 0.288 & 0.289 & 0.288 & 0.996\\
\cline{2-6}
 & \rule{0pt}{3.0ex}Global mean and spread
 & $0.292^{+0.010}_{-0.006}$
 & $0.293^{+0.010}_{-0.006}$
 & $0.292^{+0.010}_{-0.006}$
 & $0.9960^{+0.0002}_{-0.0001}$\\
\hline
$B_c\to\eta_c$ & \mbox{Log-Mix-P}
 & 0.292 & 0.293 & 0.292 & 0.997\\
 & \mbox{Lin-Mix-P}
 & 0.309 & 0.310 & 0.310 & 0.997\\
 & \mbox{Lin-Mix-NP}
 & 0.309 & 0.310 & 0.310 & 0.997\\
 & \mbox{HO-1S-P}
 & 0.307 & 0.308 & 0.307 & 0.997\\
 & \mbox{Lin-1S-P}
 & 0.327 & 0.328 & 0.328 & 0.997\\
\cline{2-6}
 & \rule{0pt}{3.0ex}Global mean and spread
 & $0.309^{+0.018}_{-0.017}$
 & $0.310^{+0.018}_{-0.017}$
 & $0.309^{+0.018}_{-0.017}$
 & $0.9968^{+0.0004}_{-0.0003}$\\
\hline
$B_c\to B$ & \mbox{Log-Mix-P}
 & -- & -- & -- & 0.950\\
 & \mbox{Lin-Mix-P}
 & -- & -- & -- & 0.953\\
 & \mbox{Lin-Mix-NP}
 & -- & -- & -- & 0.954\\
 & \mbox{HO-1S-P}
 & -- & -- & -- & 0.957\\
 & \mbox{Lin-1S-P}
 & -- & -- & -- & 0.953\\
\cline{2-6}
 & \rule{0pt}{3.0ex}Global mean and spread
 & -- & -- & --
 & $0.953^{+0.004}_{-0.003}$\\
\hline
$B_c\to B_s$ & \mbox{Log-Mix-P}
 & -- & -- & -- & 0.940\\
 & \mbox{Lin-Mix-P}
 & -- & -- & -- & 0.943\\
 & \mbox{Lin-Mix-NP}
 & -- & -- & -- & 0.944\\
 & \mbox{HO-1S-P}
 & -- & -- & -- & 0.946\\
 & \mbox{Lin-1S-P}
 & -- & -- & -- & 0.943\\
\cline{2-6}
 & \rule{0pt}{3.0ex}Global mean and spread
 & -- & -- & --
 & $0.943^{+0.003}_{-0.003}$\\
\hline\hline
\end{tabular}
\end{table*}
Table~\ref{tab:ratios} summarizes the integrated lepton-flavor
ratios for the five potential-model implementations.
The predicted ranges are $\RDRange$ for $R(D)$,
$\RDsRange$ for $R(D_s)$, and $\RetacRange$ for $R(\eta_c)$.
The $R(D)$ and $R(D_s)$ ranges largely overlap, whereas
$R(\eta_c)$ extends to higher values and exhibits a wider
model spread.

The equal-weight means and model spreads for $R(D)$ and
$R(D_s)$ are
\begin{equation}
R(D)=0.288^{+0.012}_{-0.007},
\qquad
R(D_s)=0.292^{+0.010}_{-0.006}.
\label{eq:RDsummary}
\end{equation}
The \mbox{Lin-Mix-NP} implementation gives the largest
ratios in both channels.

For $B\to D$, $B_s\to D_s$, and $B_c\to\eta_c$,
the ratios $R_{\tau/e}$ and $R_{\tau/\mu}$ are nearly
identical, with $R_{\mu/e}\simeq0.996$ for
$B_{(s)}\to D_{(s)}$ and $R_{\mu/e}\simeq0.997$
for $B_c\to\eta_c$.
Since $R_{\tau/\mu}=R_{\tau/e}/R_{\mu/e}$,
$R_{\tau/\mu}$ exceeds $R_{\tau/e}$ by less than $0.5\%$.
The light-lepton-averaged ratio $R$ lies between these
two values.

A more pronounced electron--muon difference appears
in $B_c\to B$ and $B_c\to B_s$.
Their $R_{\mu/e}$ values span $0.950$--$0.957$ and
$0.940$--$0.946$, respectively, with means of
$0.953$ and $0.943$.
The muon widths are thus suppressed relative to the
electron widths by approximately $5\%$ and $6\%$,
respectively, reflecting the greater importance of the
muon mass in the restricted phase space of these transitions.
These ratios exhibit much smaller relative model spreads
than the corresponding branching fractions.

\begin{figure*}[t]
\centering
\includegraphics[width=0.82\textwidth]{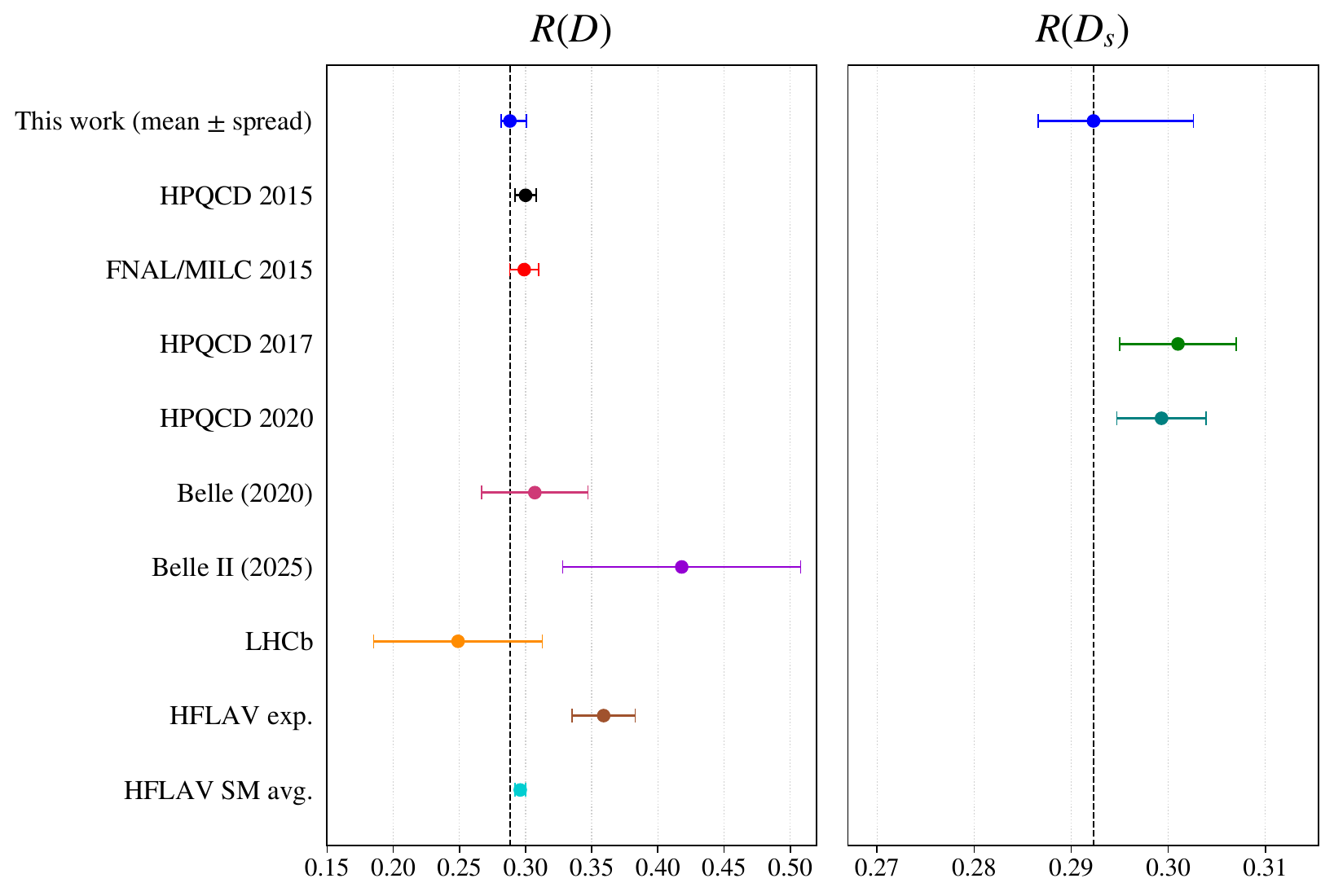}
\caption{Comparison of our five-model predictions for $R(D)$ and
$R(D_s)$ with experimental
measurements~\cite{Belle2020RD,Belle2025,LHCb2024},
the HFLAV ICHEP 2026 experimental and SM averages~\cite{HFLAV2026},
and predictions based on lattice
QCD~\cite{HPQCD2015,FNALMILC2015,HPQCD2017,McLean2020}.}
\label{fig:ratioComparison}
\end{figure*}
Figure~\ref{fig:ratioComparison} shows our predictions
for $R(D)$ and $R(D_s)$ alongside experimental measurements
and the HFLAV experimental average for $R(D)$, as well as
theoretical results for both ratios.
The blue points and asymmetric bars denote the five-model
means and spreads, while the vertical dashed lines mark
the corresponding means.
The experimental uncertainties quoted below combine
statistical and systematic contributions in quadrature,
separately for the upper and lower uncertainties.

Belle obtained $R(D)=0.307\pm0.040$ using semileptonic
tagging and a simultaneous fit to charged and neutral
$B$ decay channels with an isospin
constraint~\cite{Belle2020RD}.
For the charged-$D$ channel, Belle II and LHCb report
$R(D^+)=0.418^{+0.090}_{-0.092}$ and $0.249\pm0.064$,
respectively~\cite{Belle2025,LHCb2024}.
The figure displays the symmetrized Belle II uncertainty
adopted by HFLAV~\cite{HFLAV2026}.
Belle and Belle II use the electron--muon average in the
denominator, whereas LHCb uses the muon mode.
Our $R(D)$ range lies entirely within the quoted
one-standard-deviation interval of the Belle result
but below that of Belle II, while our $R_{\tau/\mu}(D)$
range lies entirely within the corresponding LHCb interval.

The HFLAV ICHEP 2026 update gives the nominal experimental
average $R(D)=0.359\pm0.024$~\cite{HFLAV2026}.
This average excludes the preliminary BaBar result.
Our five-model range lies below $0.335$, the lower bound
of the quoted one-standard-deviation interval for this
average.
For comparison with theory, HFLAV quotes the arithmetic
average of SM predictions, $R(D)=0.296\pm0.004$.
Its central value lies within our five-model range.
The experimental and SM averages differ by approximately
$2.6$ standard deviations for $R(D)$ alone~\cite{HFLAV2026}.

For $R(D)$, the predictions based on lattice QCD are
$0.300(8)$ from HPQCD~\cite{HPQCD2015} and
$0.299(11)$ from Fermilab/MILC~\cite{FNALMILC2015}.
For $R(D_s)$, HPQCD obtained $0.301(6)$ in
2017~\cite{HPQCD2017} and $0.2993(46)$ in
2020~\cite{McLean2020}.
All four central values lie above the corresponding
five-model means and within the respective model ranges.
The $R(D_s)$ comparison complements the differential
distributions in Fig.~\ref{fig:BsDsDiffComp}.

In their nominal heavy-quark expansion fit to lattice data,
Bordone \textit{et al.} obtained
$R_{\tau/\mu}(D)=0.3022(36)$ and
$R_{\tau/\mu}(D_s)=0.2990(31)$~\cite{Bordone2025},
using the muon mode in the denominator.
Their central value for $R_{\tau/\mu}(D)$ lies slightly
above the upper endpoint of our five-model range,
while that for $R_{\tau/\mu}(D_s)$ lies within the
corresponding range.

For $B_c\to\eta_c$, our five-model predictions for
$R_{\tau/\mu}$ span $0.293$--$0.328$, with a mean of $0.310$.
The mean matches the quoted central value of
$R_{\tau/\mu}(\eta_c)=0.310\pm0.042$ obtained by
Biswas \textit{et al.}~\cite{Biswas2025}.
Their analysis used lattice $B_c\to J/\psi$ form factors
and heavy-quark spin symmetry relations with
symmetry-breaking corrections.

\section{Summary}\label{SUM}

We have applied the fBT-LFQM formulation of
Ref.~\cite{BTChoi} to a comparative analysis of the
semileptonic $B_{(s)}\to D_{(s)}$ and
$B_c\to(\eta_c,B,B_s)$ transitions.
The five potential-model implementations use the fixed
spectroscopy-based inputs of Table~\ref{input} without
refitting to decay data.
The comparison therefore probes the combined effects
of confinement, radial mixing, hyperfine treatment,
and the associated constituent parameters within a
common current prescription.

The on-shell constituent kernels and internal Lorentz
projections use the free-pair invariant masses, while
physical meson masses enter the external kinematics
and the definition of $f_0$.
The independently integrated $(+,\perp)$ and $(+,-)$
current-component extractions yield the same $f_-(q^2)$
within numerical accuracy for every implementation
and transition.
The model dependence of $f_-$ remains pronounced,
particularly in the charm-changing $B_c$ channels.
Direct timelike continuation provides the form factors
throughout the physical decay intervals, including zero
recoil, without a form-factor fit or numerical extrapolation.
The auxiliary parametrization in Appendix~\ref{app:auxiliary}
provides a compact representation of these results and
is not used to calculate the decay observables.

For $B_{(s)}\to D_{(s)}$, all five predictions for the
zero-recoil normalizations $\mathcal G(1)$ lie below
the quoted HPQCD one-standard-deviation intervals,
whereas the HPQCD central slopes lie within our
model ranges~\cite{HPQCD2015,HPQCD2017}.
The central slopes from the BaBar and Belle II CLN fits
also lie within our $B\to D$
range~\cite{BaBar2024,BelleII2025Vcb}.

For $B_c\to B$ and $B_c\to B_s$, the HPQCD
maximum-recoil central values $f_+(0)=f_0(0)$ lie
within our model ranges.
At zero recoil, the lattice $f_+$ central value lies
within the $B_c\to B_s$ range.
For $B_c\to B$, the central value lies above the
model range, although the quoted uncertainty interval
partially overlaps it.
All five predictions for $f_0(q^2_{\max})$ lie below
the corresponding lattice one-standard-deviation
intervals in both channels~\cite{Cooper2020}.
The reduced widths, with CKM and electroweak factors
removed, have model ranges containing the lattice
central values.
Because the electron-mode widths predominantly probe
$|f_+|^2$, this overlap does not resolve the scalar
normalization differences at zero recoil.

The central values of the Belle II light-lepton branching
fractions for $B^0\to D^-$ and $B^+\to\bar D^0$
lie within the respective five-model
ranges~\cite{BelleII2025Vcb}.
For $B^0\to D^-\tau^+\nu_\tau$, all five predictions
lie below the quoted PDG one-standard-deviation
interval~\cite{PDG2024}.

The equal-weight means and model spreads of the
integrated lepton-flavor ratios are
$R(D)=0.288^{+0.012}_{-0.007}$,
$R(D_s)=0.292^{+0.010}_{-0.006}$, and
$R(\eta_c)=0.309^{+0.018}_{-0.017}$.
Our $R(D)$ range lies within the Belle
one-standard-deviation interval but below those of
Belle II and the nominal HFLAV ICHEP 2026
average~\cite{Belle2020RD,Belle2025,HFLAV2026}.
The corresponding $R_{\tau/\mu}(D)$ range lies within
the LHCb one-standard-deviation interval~\cite{LHCb2024}.
The $R(D)$ and $R(D_s)$ model ranges contain the
respective lattice central
values~\cite{HPQCD2015,FNALMILC2015,HPQCD2017,McLean2020}.
For the narrow $B_c\to B$ and $B_c\to B_s$ decay
intervals, the muon widths are about $5\%$ and $6\%$
smaller than the electron widths, respectively,
and the $\tau$ modes are kinematically forbidden.

The comparison identifies both model dependence and
discrepancies common to all five implementations.
Agreement in integrated ratios or light-lepton widths
need not imply agreement in the form-factor normalizations
at individual recoil points.
A precision CKM analysis would require correlated
uncertainty estimates and control of corrections beyond
the present constituent approximation.
Extensions to pseudoscalar-to-vector transitions will
also require independent current-component and polarization
consistency checks.

\begin{acknowledgments}
This work was supported by the National Research Foundation of Korea (NRF) grant funded by the Korea government
under Grant No. RS-2026-25475296. B.P. acknowledges the Department of Physics Education, Teachers College, Kyungpook National University for the postdoctoral fellowship during which the entire work was carried out, as well as financial support from the APOSTD Fellowship of the Generalitat Valenciana, Spain (File No. CIAPOS/2024/454).
\end{acknowledgments}
\appendix
\section{Covariant BS triangle and the fBT-LFQM}
\label{app:bs-bt}

\begin{figure*}[!ht]
\centering
\includegraphics[width=0.76\textwidth]{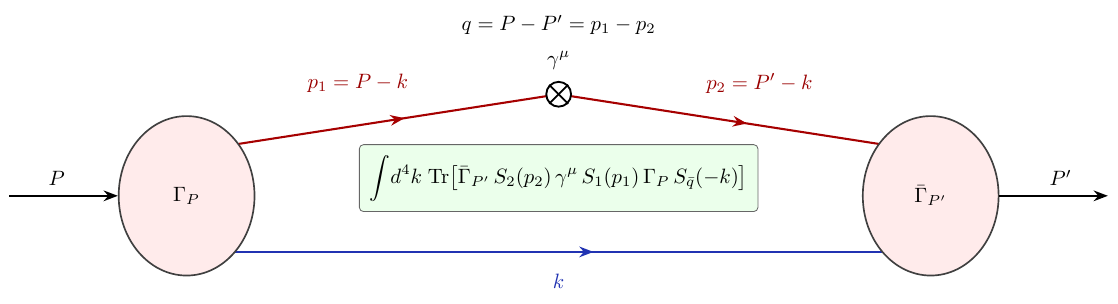}
\caption{Covariant BS triangle for
$P(q_1\bar q)\to P'(q_2\bar q)$.
The blobs denote the initial BS vertex $\Gamma_P$ and the
conjugate final vertex $\bar\Gamma_{P'}$.
The symbol $\otimes$ marks the vector-current insertion
$\gamma^\mu$, and the arrows indicate the assigned momentum flow.
The boxed expression summarizes the loop structure, with overall
normalization and model-dependent regularization factors omitted.}
\label{fig:bs-triangle}
\end{figure*}

Figure~\ref{fig:bs-triangle} shows the covariant BS
triangle with the same local vector-current vertex
$\gamma^\mu$ as in Fig.~\ref{fig:helicity-kernel}.
With spectator momentum $k$, the active-quark momenta
are $p_1=P-k$ and $p_2=P'-k$.
This routing conserves four-momentum at both meson
vertices for every loop momentum $k$.

The constituent propagator for momentum $r$ directed
along fermion flow is
\begin{equation}
S_f(r)=\frac{i(\slashed r+m_f)}{r^2-m_f^2+i0},
\label{eq:bs-fermion-propagator}
\end{equation}
where $m_f$ is the constituent mass and
$\slashed r\equiv\gamma^\mu r_\mu$.
The spectator propagator is $S_{\bar q}(-k)$ because
$k$ is opposite to fermion flow.
The internal lines are generally off shell, allowing
the routing to accommodate the physical external LF
energies.
The full BS amplitude uses the physical meson masses
$M$ and $M'$ in its Lorentz projection~\cite{BTChoi}.

For the fBT overlap, the corresponding free-pair transfer
is $q_{\rm on}=P_{\rm on}-P'_{\rm on}$.
In the DYW frame, Eqs.~\eqref{eq:externalmomenta}
and~\eqref{eq:onprefactors} give
\begin{equation}
q^- - q^-_{\rm on}
=\frac{(M^2-M_0^2)-(M'^2-M_0^{\prime2})}{P^+}.
\label{eq:bs-bt-transfer-difference}
\end{equation}
Although $q_{\rm on}^2=q^2$, the two transfers generally
have different minus components.
Unlike the BS loop, the fBT overlap contains no off-shell
propagator terms that can accommodate this difference.
Its internal Lorentz projection uses the same free-pair
kinematics as the on-shell spinor kernel.

The LF decomposition of the BS loop separates the fermion
propagators into on-shell and instantaneous parts.
A finite zero-mode contribution can survive from the
nonvalence endpoint as $q^+\to0^+$.
For the same on-shell kernel and LFWFs, the integrated
identity in Eq.~\eqref{eq:fBTcomponents} indirectly
identifies the difference between the fBT and
physical-mass minus-current extractions of $f_-$ as
the effective zero-mode correction missing from the latter.
This identification is made at the form-factor level
and does not equate the correction with the BS endpoint
operator~\cite{BTChoi}.

\section{Auxiliary parametrization of the form factors}
\label{app:auxiliary}

The predictions in Sec.~\ref{sec:results} are obtained from
direct fBT overlap calculations without assuming a parametric
form for the form factors.
To represent their timelike dependence compactly, we provide
the auxiliary rational form
\begin{equation}
f_i^{\rm aux}(q^2)
=\frac{f_+(0)}{1-b_i q^2/M^2+c_i(q^2/M^2)^2},
\qquad i=+,0,
\label{eq:auxfit}
\end{equation}
for each transition and potential-model implementation.
Here $M$ is the physical initial-meson mass, and
$0\le q^2\le q^2_{\max}=(M-M')^2$.
The common normalization $f_+(0)=f_0(0)$ is fixed by the
unrounded direct-overlap value, with the corresponding
displayed values listed in Table~\ref{tab:FFendpoint}.

Substituting Eq.~\eqref{eq:auxfit} into Eq.~\eqref{eq:f0def}
and writing $z=q^2/M^2$ and $D_i(z)=1-b_i z+c_i z^2$,
we obtain
\begin{equation}
f_-^{\rm aux}(q^2)
=\frac{M^2-M'^2}{M^2}f_+(0)\,
\frac{(b_0-b_+)+(c_+-c_0)z}{D_0(z)D_+(z)}.
\label{eq:aux-fminus}
\end{equation}
At $q^2=0$, it reduces to
$f_-^{\rm aux}(0)=(M^2-M'^2)f_+(0)(b_0-b_+)/M^2$.
Requiring this value to equal the directly calculated
$f_-(0)$ gives
\begin{equation}
b_0-b_+
=\frac{M^2 f_-(0)}{(M^2-M'^2)f_+(0)}.
\label{eq:fitslope}
\end{equation}
With $b_0$ fixed by Eq.~\eqref{eq:fitslope}, the three
independent coefficients $b_+$, $c_+$, and $c_0$ are
determined by a joint least-squares fit that minimizes
the sum of squared relative deviations,
\begin{equation}
\mathcal S(b_+,c_+,c_0)
=\sum_k\sum_{i=+,0}
\left[
\frac{f_i^{\rm aux}(q_k^2)}{f_i(q_k^2)}-1
\right]^2,
\label{eq:auxchi2}
\end{equation}
where $f_i(q_k^2)$ are the direct-overlap results
interpolated at uniformly spaced values $q_k^2$ over
$0\le q^2\le q^2_{\max}$.

Table~\ref{tab:parameterization_reduced} lists the fitted
coefficients and the maximum relative deviations
$\epsilon_+$ and $\epsilon_0$, expressed as percentages.
Here $\epsilon_i=\max_j
|f_i^{\rm aux}(q_j^2)/f_i(q_j^2)-1|$ for $i=+,0$,
where $f_i(q_j^2)$ denotes the direct-overlap result and
the maximum is taken over the directly calculated
momentum-transfer values $q_j^2$, including both endpoints.
The largest reported deviation is approximately \AuxMaxError.
These deviations quantify the accuracy of the auxiliary
representation and are distinct from numerical integration
errors and the spread among potential models.
For the narrow $B_c\to B_{(s)}$ intervals, the normalization
of $q^2$ by $M^2$ can lead to numerically large
dimensionless coefficients.

The coefficients could instead be fixed locally by
$b_i=M^2f_i'(0)/f_i(0)$ and
$c_i=b_i^2-M^4f_i''(0)/[2f_i(0)]$,
with primes denoting derivatives with respect to $q^2$.
This choice reproduces the Taylor expansion of $f_i$
through second order in $q^2$ and
automatically satisfies Eq.~\eqref{eq:fitslope}.
However, this local matching does not control the
approximation accuracy over the full physical interval.
For $B_c\to\eta_c$ with the \mbox{Lin-1S-P}
implementation, for example, it gives a larger deviation
in $f_+$ at zero recoil than the fitted representation.
We therefore use the fitted coefficients.

\begin{table*}[!htb]
\centering\small

\caption{Coefficients of the auxiliary rational representation
in Eq.~\eqref{eq:auxfit}, obtained from a joint least-squares
fit with the common direct-overlap normalization and the
constraint in Eq.~\eqref{eq:fitslope}.
The errors $\epsilon_+$ and $\epsilon_0$ are the maximum
relative deviations (in percent) at the directly calculated
momentum-transfer values, including both endpoints.}
\label{tab:parameterization_reduced}
\setlength{\tabcolsep}{13pt}
\renewcommand{\arraystretch}{1.0}
\begin{tabular}{llrrrrrr}
\hline\hline
Transition & Model & $b_+$ & $c_+$ & $b_0$ & $c_0$ & $\epsilon_+\,(\%)$ & $\epsilon_0\,(\%)$\\\hline
$B\to D$ & \mbox{Log-Mix-P} & 0.9850 & 0.1836 & 0.4688 & -0.0583 & 0.014 & 0.007\\
 & \mbox{Lin-Mix-P} & 1.0013 & 0.2098 & 0.4650 & -0.0330 & 0.007 & 0.003\\
 & \mbox{Lin-Mix-NP} & 1.1336 & 0.2327 & 0.6105 & -0.0565 & 0.038 & 0.037\\
 & \mbox{HO-1S-P} & 1.0075 & 0.2419 & 0.4841 & -0.0118 & 0.003 & 0.004\\
 & \mbox{Lin-1S-P} & 0.9557 & 0.2051 & 0.4289 & -0.0131 & 0.002 & 0.004\\
\hline
$B_s\to D_s$ & \mbox{Log-Mix-P} & 1.0637 & 0.2523 & 0.5252 & -0.0214 & 0.010 & 0.007\\
 & \mbox{Lin-Mix-P} & 1.1296 & 0.3296 & 0.5728 & 0.0290 & 0.031 & 0.019\\
 & \mbox{Lin-Mix-NP} & 1.2041 & 0.2560 & 0.6690 & -0.0838 & 0.049 & 0.025\\
 & \mbox{HO-1S-P} & 1.1810 & 0.4075 & 0.6481 & 0.0743 & 0.056 & 0.019\\
 & \mbox{Lin-1S-P} & 1.1181 & 0.3516 & 0.5710 & 0.0603 & 0.041 & 0.022\\
\hline
$B_c\to\eta_c$ & \mbox{Log-Mix-P} & 2.0500 & 1.3920 & 1.3513 & 0.6175 & 0.199 & 0.067\\
 & \mbox{Lin-Mix-P} & 2.3553 & 1.9073 & 1.6562 & 0.9430 & 0.327 & 0.112\\
 & \mbox{Lin-Mix-NP} & 2.1751 & 1.1958 & 1.5084 & 0.3670 & 0.257 & 0.104\\
 & \mbox{HO-1S-P} & 2.1379 & 1.5715 & 1.5573 & 0.7587 & 0.218 & 0.074\\
 & \mbox{Lin-1S-P} & 2.6743 & 2.4825 & 1.9595 & 1.3377 & 0.522 & 0.180\\
\hline
$B_c\to B$ & \mbox{Log-Mix-P} & 13.0991 & 67.8511 & 4.7450 & 104.7189 & 0.161 & 0.399\\
 & \mbox{Lin-Mix-P} & 15.7918 & 98.0230 & 7.2041 & 105.9149 & 0.172 & 0.326\\
 & \mbox{Lin-Mix-NP} & 14.7422 & 70.8026 & 7.7219 & 82.1056 & 0.173 & 0.399\\
 & \mbox{HO-1S-P} & 15.5910 & 88.0189 & 10.1892 & 65.2203 & 0.080 & 0.064\\
 & \mbox{Lin-1S-P} & 17.3967 & 122.0080 & 7.7568 & 111.6285 & 0.197 & 0.288\\
\hline
$B_c\to B_s$ & \mbox{Log-Mix-P} & 11.9718 & 54.9324 & 4.9104 & 81.6931 & 0.064 & 0.146\\
 & \mbox{Lin-Mix-P} & 14.3623 & 80.0914 & 7.3476 & 80.0205 & 0.062 & 0.099\\
 & \mbox{Lin-Mix-NP} & 14.1043 & 60.6501 & 8.1486 & 76.6771 & 0.087 & 0.194\\
 & \mbox{HO-1S-P} & 13.5703 & 66.5204 & 9.6068 & 49.1621 & 0.025 & 0.013\\
 & \mbox{Lin-1S-P} & 15.9487 & 102.8526 & 7.8592 & 85.6387 & 0.072 & 0.083\\
\hline\hline\end{tabular}
\end{table*}

\newpage
\bibliography{reference}
\end{document}